\documentclass{article}

\PassOptionsToPackage{numbers,compress}{natbib}

\usepackage[main, final]{neurips_2026}

\usepackage[utf8]{inputenc} 
\usepackage[T1]{fontenc}    
\usepackage{hyperref}       
\usepackage{url}            
\usepackage{booktabs}       
\usepackage{amsfonts}       
\usepackage{nicefrac}       
\usepackage{microtype}      
\usepackage{xcolor}         

\usepackage{wrapfig}
\usepackage{placeins}
\usepackage{multirow} 
\usepackage{amsmath,amssymb,amsthm,mathtools,bm}
\usepackage{graphicx}
\usepackage{subcaption}
\usepackage{tikz}
\usepackage{colortbl}
\usetikzlibrary{calc, arrows.meta}

\newtheorem{proposition}{Proposition}
\newtheorem{lemma}{Lemma}
\newtheorem{corollary}{Corollary}
\theoremstyle{definition}

\theoremstyle{remark}

\newcommand{\Tr}{\operatorname{Tr}}

\newcommand{\loc}{\mathrm{loc}}

\newcommand{\cB}{\mathcal{B}}

\newcommand{\cS}{\mathcal{S}}
\newcommand{\cV}{\mathcal{V}}
\newcommand{\cW}{\mathcal{W}}
\newcommand{\RR}{\mathbb{R}}

\newcommand{\ket}[1]{|#1\rangle}
\newcommand{\bra}[1]{\langle #1|}

\newcommand{\PE}{\mathrm{PE}}

\newcommand{\sqtabsize}{\small}
\newcommand{\sqtabsetup}{%
  \sqtabsize
  \setlength{\tabcolsep}{2.0pt}%
  \renewcommand{\arraystretch}{1.05}%
}
\newcommand{\best}[1]{\textbf{#1}}

\providecommand{\vMemOffCello}{36}
\providecommand{\vMemOffAstronaut}{388}
\providecommand{\vMemOffCamera}{291}
\providecommand{\vMemOffCoffee}{149}

\providecommand{\vNeRFParamRatio}{1.54}
\providecommand{\vShotGapPct}{59.95}

\title{Spectral Quantum Memory for Implicit Neural Representations}

\author{%
  Daniel A. Moroz-Broitman \\
  Department of Computer Science\\
  Ben Gurion University\\
  \texttt{morozd@post.bgu.ac.il}
  \And
  Eliya Nachmani \\
  Department of Electrical Engineering \\
  Ben Gurion University \\
  \texttt{eliyanac@bgu.ac.il}
}

\begin{document}
\maketitle

\begin{abstract}
Implicit neural representations (INRs) model signals as continuous functions from coordinates to values, but their performance depends strongly on how spectral capacity is allocated. Yet most Quantum INR (QINR) architectures treat spectral components as independent outputs of separate quantum blocks. We introduce the \emph{Spectral Quantum Memory Model} (SQMM), a memory coupled quantum INR in which sequential hybrid data reuploading layers (bands) communicate through a shared coherent quantum memory register. Each band reads the memory, computes on its own coordinate upload, and writes part of its state back, so later bands can build on the computation of earlier ones. When memory read and write layers are input independent, we prove that the memory state forms an operator valued Fourier series whose support is contained in the Minkowski sum of the initial memory spectrum and the local band spectra. Downstream bands then condition their Fourier coefficients on the spectral content stored by earlier bands, and the memory coupling creates no new primitive frequencies. This theory motivates a \emph{spectral memory alignment} loss that encourages memory to encode the spectrum expressed by the deployed prefix of the model. Across audio and image reconstruction benchmarks, SQMM improves over compact QINR and INR baselines. Code is provided in: https://github.com/Kiiwii10/SQMM
\end{abstract}

\section{Introduction}\label{sec:intro}

\begin{figure}[t]
  \centering
  \includegraphics[width=0.9\linewidth]{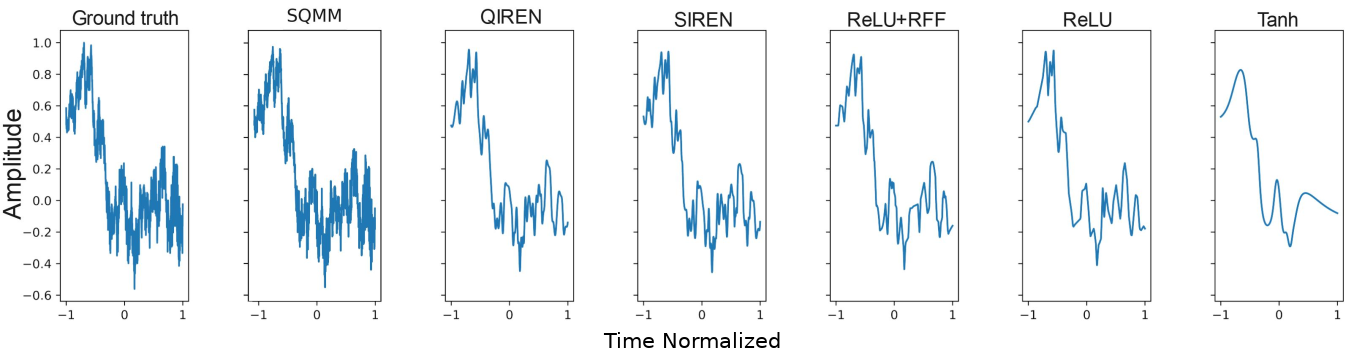}
  \caption{Sound representation results adapted from QIREN~\citep{zhao2024quantumimplicitneuralrepresentations}, showing normalized amplitude over normalized time.}
  \label{fig:cello_wave}
\end{figure}

Implicit neural representations (INRs) model signals as continuous coordinate to value functions, enabling compact representations for images, audio (Figure~\ref{fig:cello_wave}), video, geometry, radiance fields, and physical signals \citep{xie2022neuralfields,park2019deepsdf,mildenhall2020nerf,sitzmann2020siren,dupont2021coin,tancik2020fourier}. Their main bottleneck is spectral. Standard MLPs learn low frequency structure before high frequency detail \citep{rahaman2019spectral,xu2019frequency}, so INR performance depends on how frequency content is exposed, selected, and composed \citep{benbarka2022seeing}. This has motivated Fourier features \citep{tancik2020fourier}, periodic activations \citep{sitzmann2020siren}, multiplicative filters \citep{fathony2021multiplicative}, band limited and multiscale coordinate networks \citep{Lindell2022BACON,Shekarforoush2022rMFN}, wavelet INRs \citep{saragadam2023wire}, and adaptive periodic bases \citep{liu2023finerflexiblespectralbiastuning,mehrabian2025fkan}. These works show that spectral organization is central to INR capacity.

Quantum implicit neural representations (QINRs) make this spectral structure explicit because data reuploading circuits compute finite Fourier series. Data reuploading repeatedly injects classical inputs into variational quantum circuits \citep{mitarai2018quantum,PerezSalinas2020Data,VidalTheis2020input}. Their accessible frequencies are determined by encoding gates, while trainable circuit parameters determine the coefficients \citep{schuld2021effect,Casas2023Multidimensional,yu2022power,WiedmannPeriyasamyScherer2024fourier,BarthePerezSalinas2024gradients,MhiriEtAl2025constrained}. QIREN formulates QINRs as quantum generalizations of Fourier neural networks \citep{zhao2024quantumimplicitneuralrepresentations}, and QVF extends QINRs toward quantum centric visual fields through neural amplitude encoding and projective measurement \citep{Wang2025QVF}. Quantum recurrent models and quantum LSTMs use hidden quantum states or variational cells as memory mechanisms \citep{Bausch2020rqnn,chen2022quantum}. Existing QINRs, however, do not characterize how memory qubits share spectral information across coordinate frequency modules.

In existing multiband QINRs, the bands operate largely independently, and what one band computes reaches the others only through the classical decoder. A shared memory lets information produced by earlier bands influence later ones. This paper introduces the Spectral Quantum Memory Model (SQMM), a memory coupled QINR built from sequential data reuploading circuits organized into hybrid quantum and classical bands. Each band reads a shared unmeasured memory register, performs local spectral computation, and writes back before the next band. SQMM therefore propagates coherent spectral context across bands, allowing later bands to condition coefficient allocation on earlier spectral computation instead of relearning each band independently.

When memory read and write layers are independent of the input coordinate, each memory coupled band remains in the data reuploading model class. The reduced memory state is an operator valued Fourier series whose support is contained in the Minkowski sum of the initial memory spectrum and the band spectra. Downstream readout coefficients arise from convolutions between the memory spectrum and local band spectra. Thus memory does not introduce arbitrary primitive frequencies. It preserves the Fourier properties of QINRs while expanding spectral context between bands.

This view motivates a spectral memory alignment loss. The loss aligns memory with an estimate of the accumulated spectral occupancy of the deployed prefix contribution. The supervised loss trains the predictor, while the alignment loss calibrates memory as a readable spectral summary of the computation so far. Downstream bands can still reuse, refine, or correct upstream frequencies.

The contributions are fourfold. First, SQMM introduces a memory coupled QINR that propagates coherent spectral context across sequential data reuploading bands. Second, the theory proves that memory preserves the Fourier properties of data reuploading circuits, since the memory state remains an operator valued Fourier series whose support is contained in the Minkowski sum of the band spectra. Third, a spectral memory alignment loss trains memory to summarize the deployed prefix spectrum. Fourth, SQMM improves on the reconstruction results of quantum and classical INR baselines on audio and image representation tasks under small parameter budgets.

\section{Related Work}\label{sec:related}

\textbf{Implicit neural representations and spectral coordinate networks.}
Coordinate based neural representations originate in compositional pattern producing networks \citep{stanley2007compositional} and now support continuous representation of signals, shapes, scenes, and fields \citep{xie2022neuralfields}, including shape templates \citep{genova2019learning}, continuous shape distributions \citep{zhong2019reconstructing}, signed distance functions \citep{park2019deepsdf}, scene representation networks \citep{sitzmann2019scene}, local implicit grids \citep{jiang2020local}, neural radiance fields \citep{mildenhall2020nerf}, image generation and compression \citep{skorokhodov2021adversarial,anokhin2021image,dupont2021coin,chan2021pi,orel2022stylesdf}, time indexed modeling \citep{woo2023learning}, and scientific modeling \citep{huang2022universal}. Their central limitation is spectral bias, since standard networks tend to fit low frequencies before high frequency detail \citep{rahaman2019spectral,xu2019frequency,basri2020frequency,tancik2020fourier}. This has motivated spectral INR designs based on Fourier features \citep{tancik2020fourier}, sinusoidal and nonstandard activations \citep{parascandolo2016taming,sitzmann2020siren,ramasinghe2022beyond}, multiplicative filters \citep{fathony2021multiplicative}, band limited and multiscale networks \citep{Lindell2022BACON,Shekarforoush2022rMFN}, wavelet INRs \citep{saragadam2023wire}, adaptive periodic bases \citep{liu2023finerflexiblespectralbiastuning,mehrabian2025fkan}, Fourier reparameterized training \citep{shi2024fourier}, and, more recently, content aware Fourier Chebyshev encodings (CAFE)~\citep{cafe2026}, multi scale sine activations (MSA-INR)~\citep{msainr2026}, frequency and locality aware representations (FLAIR)~\citep{flair2026}, robust frequency tuning for sinusoidal networks (TUNER)~\citep{tuner2025}, variance informed initialization (VI3NR)~\citep{vi3nr2025}, and inductive gradient adjustment against spectral bias (IGA)~\citep{iga2025}. These methods shape the spectrum of a single classical network through its encoding, activation, initialization, or gradient, and none of them passes a state between separately computed frequency modules. SQMM follows this spectral view, but studies spectral sharing in quantum data reuploading circuits where the finite Fourier structure can be analyzed explicitly.

\textbf{Parameterized quantum models and data reuploading.}
Parameterized quantum circuits are trainable models for supervised learning, quantum feature maps, generative modeling, reinforcement learning, and hybrid quantum classical computation \citep{mcclean2016theory,mitarai2018quantum,farhi2018classification,dallaire2018quantum,havlivcek2019supervised,Benedetti2019PQCML,Sim2019Expressibility,schuld2019evaluating,bergholm2018pennylane}. Hybrid quantum neural networks combine classical processing with quantum circuits that implement trainable feature spaces, kernels, or intermediate representations \citep{schuld2020circuit,Goto2021Universal}, with applications in sequence modeling, classification, reinforcement learning, generative modeling, and architecture search \citep{chen2022quantum,li2022quantum,chen2022quan,lockwood2020reinforcement,gao2022enhancing,wang2022quantumnas}. Data reuploading repeatedly injects classical variables into variational quantum circuits and improves expressivity even with small registers \citep{PerezSalinas2020Data,VidalTheis2020input}. Angle encoded quantum models compute finite Fourier series whose accessible frequencies are determined by encoding generator spectra, while trainable parameters determine the coefficients \citep{schuld2021effect}. This Fourier view has been extended to single qubit native models \citep{yu2022power}, multidimensional Fourier series \citep{Casas2023Multidimensional}, universal approximation in quantum enhanced feature spaces \citep{Goto2021Universal}, structured and exponential encodings \citep{shin2023exponential,wach2023data}, exact realized spectra and coefficients \citep{WiedmannPeriyasamyScherer2024fourier}, coefficient constraints from encoding redundancy \citep{MhiriEtAl2025constrained}, and gradient profiles in deep reuploading circuits \citep{BarthePerezSalinas2024gradients}. SQMM extends this Fourier framework from isolated reuploading circuits to stacked memory coupled bands.

\textbf{Quantum implicit neural representations.}
The closest prior work is QIREN, which connects data reuploading circuits to INRs by treating parameterized quantum circuits as quantum analogues of Fourier feature networks \citep{zhao2024quantumimplicitneuralrepresentations}. QIREN uses repeated coordinate encodings to produce quantum Fourier features decoded by classical layers for signal representation, image super resolution, and image generation. QVF moves toward quantum centric visual fields using neural amplitude encoding and projective measurement over entangled ansatzes for 2D and 3D signals \citep{Wang2025QVF}. SQMM is complementary to both models. It retains the data reuploading Fourier view of QIREN but couples its bands through a shared coherent memory, and it differs from QVF by emphasizing spectral communication across reuploading bands rather than amplitude encoding or direct visual field measurement. Prior Fourier analyses of data reuploading circuits characterize circuits with fixed incoming quantum states~\citep{schuld2021effect,yu2022power,Casas2023Multidimensional}. QIREN connects that theory to INRs, but it measures each quantum block and passes a classical feature vector to the next hybrid layer, so its analysis characterizes the Fourier form of an individual circuit in its immediate input. SQMM instead keeps a coherent memory state across sequential bands, and our analysis describes how the operator valued spectrum of that state conditions downstream quantum features throughout the model.

\textbf{Memory, recurrence, and auxiliary spectral supervision.}
Memory and recurrence have appeared in quantum learning through recurrent quantum neural networks, quantum LSTM style architectures, and quantum reservoir computing \citep{Bausch2020rqnn,chen2022quantum,Fujii2017QRC}. Classical models also use inter stage coupling through residual or stagewise refinement, where later modules correct earlier computations \citep{HeZhangRenSun2016resnet,Friedman2001gbm}. In INRs, band limited and multiscale coordinate networks explicitly organize frequency bands \citep{Lindell2022BACON,Shekarforoush2022rMFN}, while spectral regularization and auxiliary supervision shape learned representations during training \citep{shi2024fourier,Lee2015DSN}. Quantum recurrent models use hidden quantum states as memory, but they do not characterize the coordinate frequency content of that state or its effect on downstream QINR coefficients. SQMM differs from these classical mechanisms because the coupled state is not a residual stream, recurrent hidden vector, or auxiliary prediction head. It is a quantum density matrix with operator valued Fourier coefficients that remains readable by downstream quantum bands.

\section{Background}
\label{sec:background}

We consider coordinate based signal representation. Coordinates are \(x\in\RR^d\), targets are \(y\in\RR^{d_y}\), and the model learns a continuous map \(x\mapsto y\). SQMM first maps \(x\) to a band specific uploaded variable
\begin{equation}
\gamma_i(x)=A_i\PE(x)+b_i\in\RR^{g_i},
\qquad
A_i\in\RR^{g_i\times p},
\qquad
b_i\in\RR^{g_i}.
\label{eq:upload-map}
\end{equation}
Here \(\PE:\RR^d\to\RR^p\) is a fixed positional or Fourier feature encoding~\citep{vaswani2017attention,tancik2020fourier,mildenhall2020nerf}. SQMM uses quantum data reuploading circuits inside this coordinate model. A register is a group of qubits treated as one subsystem. Band \(i\) has a local data register \(D_i\) with \(q_i\) qubits, and all bands share a memory register \(M\) with \(m\) qubits. The fixed new data state is \(\eta_{D_i}\), and the memory state after band \(i\) is a density matrix \(\sigma_M^{(i)}(x)\), namely a positive semidefinite matrix with trace one. A quantum operation is written as a channel. The band channel \(\cB_i(\gamma_i(x))\) acts on the joint register \(D_i\otimes M\). Table~\ref{tab:notation} in Appendix~\ref{app:notation} collects the notation.

We use \(\Tr[\cdot]\) for trace and \(\Tr_{D_i}[\cdot]\) for partial trace over the local data register. An observable \(O\) is a Hermitian matrix that represents a measurable quantity, and \(\Tr[O\rho]\) is its expectation value in state \(\rho\). In SQMM these expectations are used as classical features.

A finite data reuploading circuit has a finite Fourier expansion in the uploaded variable because input dependent gates are generated by finite dimensional Hermitian operators \citep{schuld2021effect,Casas2023Multidimensional,yu2022power}. For a generic uploaded variable \(z\in\RR^{g_i}\), the local upload support of band \(i\) is denoted by \(\Omega_{i,\gamma}^{\loc}\), with terms \(e^{\mathrm{i}\langle\omega_\gamma,z\rangle}\) for \(\omega_\gamma\in\Omega_{i,\gamma}^{\loc}\). After substituting \(z=\gamma_i(x)\), the analysis uses the induced coordinate support \(\Omega_{i,x}^{\loc}\). This support is exact when the upload is affine in the raw coordinate. With fixed sinusoidal positional encodings or nonlinear angle maps, as commonly used in coordinate networks and QINRs \citep{tancik2020fourier,mildenhall2020nerf,zhao2024quantumimplicitneuralrepresentations}, the composed coordinate dependence is analyzed either in the lifted encoding coordinates or through the finite grid projection used by the spectral alignment loss. Thus the theory applies to an exact finite coordinate expansion when available, and otherwise to the chosen finite truncation. For sets \(U,V\subset\RR^d\), let \(U+V=\{u+v\mid u\in U,\ v\in V\}\) denote their Minkowski sum. Since memory is a density matrix, its Fourier coefficients are matrices rather than scalar hidden features. Additional quantum notation and the upload to coordinate support conversion are given in Appendix~\ref{app:prelim}.
\section{Spectral Quantum Memory Model}
\label{sec:method}

\begin{figure}[t]
  \centering
  \resizebox{0.9\linewidth}{!}{\input{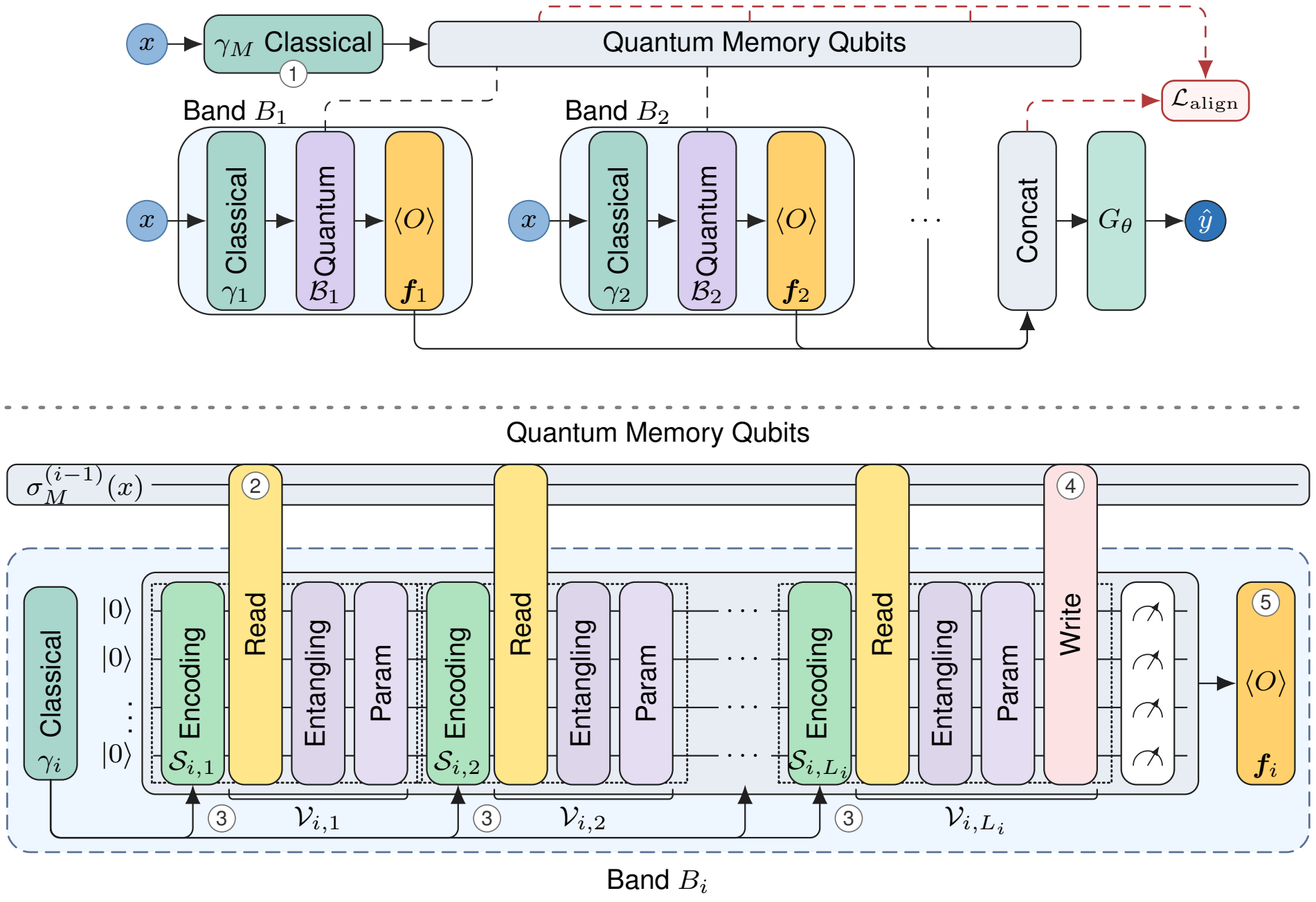}}
    \caption{SQMM architecture. Coordinates are mapped to band uploads \(\gamma_i(x)\) and to the memory upload \(\gamma_M(x)\), both of the form of Eq.~\eqref{eq:upload-map}. Sequential quantum bands share the memory register \(M\) and emit Pauli expectation features, which a classical decoder combines. Dashed red arrows mark the training only spectral alignment loss. Numbers follow the Intuition paragraph.}
  \label{fig:architecture}
\end{figure}

SQMM (Figure~\ref{fig:architecture}) is a hybrid quantum and classical INR for learning \(x\mapsto y\). It consists of \(n\) sequential data reuploading bands connected by a shared memory register. Each band emits expectation valued classical features, while the memory state is updated and passed to later bands. The memory read and write layers are input independent, so they change Fourier coefficients but do not create new primitive coordinate frequencies.

\textbf{Intuition.} Without memory, each band of a multiband QINR sees only the coordinate, and its features meet those of other bands only in the classical decoder. SQMM adds a small register of \(m\) unmeasured memory qubits that is carried from band to band. For a coordinate \(x\), the memory is first prepared from the coordinate by a one time upload \(\gamma_M(x)\)~(1). Each band then reads the memory through input independent couplings~(2), uploads its own coordinate encoding \(\gamma_i(x)\) several times~(3), and writes part of its local state back into memory~(4) before its local qubits are discarded. The band emits Pauli expectation features, and the updated memory is passed to the next band~(5). (1)-(5) refer to marks in Figure~\ref{fig:architecture}. At inference only the band features are estimated, and the memory is never measured. Section~\ref{sec:theory} shows what this pathway can carry, and Section~\ref{sec:loss} introduces a loss that trains the memory to resemble the spectrum accumulated by the preceding bands.

\subsection{Architecture}
\label{sec:architecture}

Band \(i\) starts with a new data state \(\eta_{D_i}\) and incoming memory \(\sigma_M^{(i-1)}(x)\). Its uploaded variable is \(\gamma_i(x)\) from Eq.~\eqref{eq:upload-map}. The band channel is
\begin{equation}
\cB_i(\gamma_i(x))
=
\cV_{i,L_i+1}\circ
\cS_{i,L_i}(\gamma_i(x))\circ\cV_{i,L_i}\circ\cdots\circ
\cS_{i,1}(\gamma_i(x))\circ\cV_{i,1}.
\label{eq:band-form}
\end{equation}
The layers \(\cS_{i,\ell}\) are the only input dependent layers. The layers \(\cV_{i,\ell}\) are trainable and input independent, and include local gates, entangling gates, memory read couplings, memory write couplings, and memory only processing. In the deployed circuit, \(\cV_{i,1}\) holds the memory read couplings, each upload is followed by a local entangling layer, and the closing block \(\cV_{i,L_i+1}\) holds the last local entangling layer together with the memory write couplings. Therefore memory coupling preserves the data reuploading form on \(D_i\otimes M\). The joint post band state is
\begin{equation}
\xi_i(x)
=
\cB_i(\gamma_i(x))
\left(\eta_{D_i}\otimes\sigma_M^{(i-1)}(x)\right).
\label{eq:joint-post-band-state}
\end{equation}
The updated memory is obtained by discarding the local data register,
\begin{equation}
\sigma_M^{(i)}(x)
=
\Tr_{D_i}\left[\xi_i(x)\right].
\label{eq:memory-update-main}
\end{equation}
Band \(i\) emits Pauli expectation features
\begin{equation}
\bm f_i(x)
=
\left(
\Tr[(O_{i,1}\otimes I_M)\xi_i(x)],
\ldots,
\Tr[(O_{i,d_{f,i}}\otimes I_M)\xi_i(x)]
\right)
\in\RR^{d_{f,i}}.
\label{eq:feature-readout}
\end{equation}
These features are ensemble expectations. The downstream memory is not conditioned on local measurement outcomes. Hardware estimation uses repeated circuit executions, and the readout semantics are formalized in Appendix~\ref{app:readout}. Let \(d_F=\sum_{i=1}^n d_{f,i}\), let
\begin{equation}
\bm f(x)
=
[\bm f_1(x)^\top,\ldots,\bm f_n(x)^\top]^\top
\in\RR^{d_F},
\end{equation}
and let \(G_\theta:\RR^{d_F}\to\RR^{d_y}\) be a classical decoder. The prediction is
\begin{equation}
\hat y(x)=G_\theta(\bm f(x)).
\label{eq:final-readout}
\end{equation}

\subsection{Spectral Interpretation}
\label{sec:theory}

A data reuploading circuit computes a finite Fourier series whose frequencies are fixed by its encoding gates. The result below states that this remains true for a band that receives a coordinate dependent memory state. Every frequency that leaves a band, in its features or in the memory passed on, is the sum of a frequency already present in the incoming memory and a frequency introduced by the band's own uploads. The input independent layers of the band decide how strongly each such frequency appears. The memory is therefore a channel through which later bands can reuse and reweight the spectral content of earlier bands.

The spectral analysis expands each band on the coordinate support \(\Omega_{i,x}^{\loc}\). For a fixed band \(i\), assume the incoming memory has the operator valued Fourier expansion
\begin{equation}
\sigma_M^{(i-1)}(x)
=
\sum_{\mu\in\Omega_{M,x}^{(i-1)}}
e^{\mathrm{i}\langle\mu,x\rangle}
\tau_\mu^{(i-1)},
\qquad
\tau_\mu^{(i-1)}\in\mathbb C^{2^m\times2^m}.
\label{eq:memory-fourier-main}
\end{equation}
Here \(\tau_\mu^{(i-1)}\) is the matrix coefficient of the memory state at coordinate frequency \(\mu\). Write the coordinate expansion of the band channel as
\begin{equation}
\cB_i(\gamma_i(x))
=
\sum_{\nu\in\Omega_{i,x}^{\loc}}
e^{\mathrm{i}\langle\nu,x\rangle}
\cB_{i,\nu}.
\label{eq:band-fourier-main}
\end{equation}

\begin{lemma}[Spectral propagation through memory]
\label{lem:spectral-propagation}
Under Eqs.~\eqref{eq:memory-fourier-main} and~\eqref{eq:band-fourier-main}, the outgoing memory and emitted features of band \(i\) have no coordinate frequencies outside \(\Omega_{M,x}^{(i-1)}+\Omega_{i,x}^{\loc}.\) In particular,
\begin{equation}
\Omega_{M,x}^{(i)}
\subseteq
\Omega_{M,x}^{(i-1)}+\Omega_{i,x}^{\loc},
\qquad
\Omega_{f_i,x}
\subseteq
\Omega_{M,x}^{(i-1)}+\Omega_{i,x}^{\loc}.
\label{eq:memory-spectrum-main}
\end{equation}
Input independent read, write, and memory only layers can change Fourier coefficients but cannot enlarge these supports (Proposition~\ref{prop:band-closure} and Lemma~\ref{lem:closure}).
\end{lemma}

\begin{proof} The main point is the product rule for Fourier phases. Substituting Eqs.~\eqref{eq:memory-fourier-main} and~\eqref{eq:band-fourier-main} into Eq.~\eqref{eq:memory-update-main} gives
\begin{equation}
\sigma_M^{(i)}(x)
=
\sum_{\omega}
e^{\mathrm{i}\langle\omega,x\rangle}
\tau_\omega^{(i)},
\qquad
\tau_\omega^{(i)}
=
\sum_{\mu+\nu=\omega}
\Tr_{D_i}
\left[
\cB_{i,\nu}
\left(\eta_{D_i}\otimes\tau_\mu^{(i-1)}\right)
\right].
\label{eq:memory-recursion-main}
\end{equation}
Thus an outgoing memory frequency \(\omega\) can appear only when \(\omega=\mu+\nu\), where \(\mu\) is already present in memory and \(\nu\) is introduced by the current band. The same argument gives the emitted feature spectrum. For feature component \(a\), define
\begin{equation}
\ell_{i,a,\nu}(X)
=
\Tr
\left[
(O_{i,a}\otimes I_M)
\cB_{i,\nu}
(\eta_{D_i}\otimes X)
\right].
\label{eq:feature-functional-main}
\end{equation}
This map is linear in the memory side matrix \(X\). Applying it to Eq.~\eqref{eq:memory-fourier-main} gives
\begin{equation}
f_{i,a}(x)
=
\sum_{\omega\in\Omega_{M,x}^{(i-1)}+\Omega_{i,x}^{\loc}}
c_{\omega,a}^{(i)}
e^{\mathrm{i}\langle\omega,x\rangle},
\qquad
c_{\omega,a}^{(i)}
=
\sum_{\mu+\nu=\omega}
\ell_{i,a,\nu}
\left(\tau_\mu^{(i-1)}\right).
\label{eq:output-fourier-main}
\end{equation}
This proves the support claim.
\end{proof}
The lemma shows that memory does not introduce arbitrary primitive frequencies. It lets downstream bands allocate coefficients conditional on an operator valued spectral summary of upstream computation. The one time memory upload \(\gamma_M(x)\) before the first band (Figure~\ref{fig:architecture}) gives the base support \(\Omega_{M,x}^{(0)}\), which is \(\{0\}\) when memory is initialized independently of \(x\). Applying Eq.~\eqref{eq:memory-spectrum-main} recursively bounds the memory support after band \(i\) by \(\Omega_{M,x}^{(0)}+\sum_{j=1}^{i}\Omega_{j,x}^{\loc}\) (Proposition~\ref{prop:memory-init-encoding} and Corollary~\ref{cor:global-support} in Appendix~\ref{app:proofs}).

\subsection{Spectral Memory Alignment Loss}
\label{sec:loss}

Lemma~\ref{lem:spectral-propagation} says that memory can carry a spectral summary of the prefix computation. The alignment loss makes this summary readable. It compares two spectra on a finite grid,
\begin{equation}
\Omega_{\mathrm{grid}}
=
\{\bar\omega_1,\ldots,\bar\omega_{N_\omega}\}
\subset\RR^d.
\end{equation}
Bars denote grid frequencies. Hats denote coefficients estimated from sampled coordinates.

\textbf{Feature spectrum.}
For sampled coordinates \(x_1,\ldots,x_T\), define
\begin{equation}
\Phi_{th}
=
e^{\mathrm{i}\langle\bar\omega_h,x_t\rangle},
\qquad
\widehat a(u)
=
(\Phi^\dagger\Phi+\varepsilon I_{N_\omega})^{-1}\Phi^\dagger u.
\label{eq:fourier-ls-main}
\end{equation}
All signals in this paper are sampled on regular grids. We therefore take \(\Omega_{\mathrm{grid}}\) from the discrete Fourier frequencies of the sampling grid and set \(\varepsilon=0\), so Eq.~\eqref{eq:fourier-ls-main} reduces to the discrete Fourier transform scaled by \(1/T\) and is computed with an FFT (Appendix~\ref{app:align-details}).
Applying this estimator to each feature component of band \(j\) gives \(\widehat{\bm c}_{j,h}^F=(\widehat c^F_{j,1,h},\ldots,\widehat c^F_{j,d_{f,j},h})\in\mathbb C^{d_{f,j}}\), and the prefix feature occupancy is
\begin{equation}
P_i^F(\bar\omega_h)
=
\sum_{j=1}^{i}
\|\widehat{\bm c}_{j,h}^F\|_2^2.
\label{eq:prefix-spectrum}
\end{equation}
This records which grid frequencies have been expressed by the emitted features of bands \(1,\ldots,i\), before the final decoder.

\textbf{Memory spectrum.}
The memory is probed with the unnormalized \(m\) qubit Pauli string basis \(\mathcal Q_m=\{I,X,Y,Z\}^{\otimes m}\), with \(|\mathcal Q_m|=4^m\) and \(\Tr(Q_sQ_{s'})=2^m\mathbf 1\{s=s'\}\). For each \(Q_s\in\mathcal Q_m\), the scalar memory probe is \(\chi_{i,s}^M(x)=\Tr[Q_s\sigma_M^{(i)}(x)]\). Estimating its Fourier coefficients with Eq.~\eqref{eq:fourier-ls-main} and collecting them at \(\bar\omega_h\) gives \(\widehat{\bm c}_{i,h}^M=(\widehat c^M_{i,s,h})_{Q_s\in\mathcal Q_m}\in\mathbb C^{4^m}\). The memory occupancy spectrum is
\begin{equation}
P_i^M(\bar\omega_h)
=
2^{-m}
\sum_{Q_s\in\mathcal Q_m}
|\widehat c^M_{i,s,h}|^2 .
\label{eq:memory-spectrum-loss-main}
\end{equation}
The factor \(2^{-m}\) follows from the Pauli trace normalization and cancels under the normalization of Eq.~\eqref{eq:align-normalization} up to a rescaling of \(\delta\). If the finite expansion is exact and the full Pauli basis is used, then \(P_i^M(\bar\omega_h)=\|\tau_{\bar\omega_h}^{(i)}\|_F^2\) up to finite grid estimation error. The derivation is given in Appendix~\ref{app:align-details}.

\textbf{Alignment objective.}
For any nonnegative spectrum \(R:\Omega_{\mathrm{grid}}\to\RR_{\ge0}\), define
\begin{equation}
\mathcal N_\delta R(\bar\omega_h)
=
\frac{R(\bar\omega_h)+\delta}
{\sum_{g=1}^{N_\omega}R(\bar\omega_g)+N_\omega\delta}.
\label{eq:align-normalization}
\end{equation}
With stop gradient \(\operatorname{sg}\), the alignment loss is
\begin{equation}
\mathcal L_{\mathrm{align}}
=
\sum_{i=1}^{n-1}
\sum_{h=1}^{{N_\omega}}
\left(
\mathcal N_\delta P_i^M(\bar\omega_h)
-
\operatorname{sg}
\left[
\mathcal N_\delta P_i^F(\bar\omega_h)
\right]
\right)^2.
\label{eq:align-loss}
\end{equation}
The sum stops at \(n-1\), since the final memory state is not used by a downstream band. The stop gradient keeps the prefix feature spectrum as a target and trains the memory state to match it. For supervised loss \(\ell_{sup}\), task loss \(\mathcal L_{\mathrm{task}}
=
\frac{1}{|\mathcal D|}
\sum_{(x,y)\in\mathcal D}
\ell_{\mathrm{sup}}(\hat y(x),y)\), the full training objective is
\begin{equation}
\mathcal L
=
\mathcal L_{\mathrm{task}}
+
\lambda_{\mathrm{align}}
\mathcal L_{\mathrm{align}}.
\label{eq:loss}
\end{equation}
The supervised term trains the predictor. The auxiliary term encourages the memory after each deployed prefix to store the resolved spectrum of that prefix. Frequencies outside \(\Omega_{\mathrm{grid}}\) are not matched directly and are treated only through the finite grid projection. In our experiments the loss is averaged over boundaries and grid frequencies, weighted by \(\lambda_{\mathrm{align}}\), and applied only while bands are being added (Appendix~\ref{app:align-details}).

\section{Experiments}\label{sec:exp}

We evaluate SQMM as a compact coordinate based reconstruction model for audio and images. The comparison targets the low parameter regime used by QIREN~\citep{zhao2024quantumimplicitneuralrepresentations}, which is the closest quantum INR baseline. All experiments were run on internal CPU cluster nodes with a mixed state simulator using PennyLane~\citep{bergholm2018pennylane} and PyTorch~\citep{paszke2019pytorch}. Unless stated otherwise, models were trained with Adam~\citep{kingma2015adam} for 600 epochs using learning rate \(0.03\), betas \((0.9,0.999)\), epsilon \(10^{-8}\), and supervised MSE loss.

\textbf{Datasets.} The datasets cover single signal reconstruction, dataset level audio reconstruction, and image reconstruction. The classic INR setting follows QIREN~\citep{zhao2024quantumimplicitneuralrepresentations} and uses a Bach cello excerpt together with Astronaut, Camera, and Coffee images from scikit image~\citep{van2014scikit}. Dataset level audio uses GTZAN~\citep{tzanetakis2002musical}, UrbanSound8K~\citep{salamon2014dataset}, and VCTK~\citep{yamagishi2019vctk}. To control simulation cost, each selected audio dataset uses 10 samples per class. Since only VCTK has more than 10 speakers (classes), the first 10 are chosen. CIFAR10~\citep{krizhevsky2009learning} is used for grayscale image reconstruction on 100 images. For larger images we use all 24 Kodak images~\citep{kodak1999} and 10 DIV2K images~\citep{agustsson2017div2k}, in RGB.

\textbf{Baselines.} Baselines include classical coordinate networks and quantum INR models. We compare against ReLU, Tanh, ReLU with random Fourier features (R+RFF)~\citep{tancik2020fourier}, SIREN~\citep{sitzmann2020siren}, FINER~\citep{liu2023finerflexiblespectralbiastuning}, WIRE~\citep{saragadam2023wire}, KAN~\citep{liu2025kan}, NeRF~\citep{mildenhall2020nerf}, QIREN~\citep{zhao2024quantumimplicitneuralrepresentations}, and QVF~\citep{Wang2025QVF}, and against the recent INRs CAFE~\citep{cafe2026}, MSA-INR~\citep{msainr2026}, FLAIR~\citep{flair2026}, TUNER~\citep{tuner2025}, VI3NR~\citep{vi3nr2025}, and IGA~\citep{iga2025}. Every classical baseline has one configuration for audio and one for images. Each uses the default settings of its official implementation, with network width and depth selected inside a fixed parameter budget and the learning rate selected from \([3\cdot10^{-2},10^{-3}]\). Configurations and the two baselines with adjusted settings, MSA-INR and the encoder scaled CAFE (enc.), are given in Appendix~\ref{app:config}.

\textbf{Metrics.} Metrics are reported as reconstruction error unless otherwise stated. Audio and classic INR tables report MSE with the scaling shown in each caption. Audio analysis also reports log spectral distance. CIFAR10 reports MSE, SSIM~\citep{wang2004image}, and LPIPS~\citep{zhang2018unreasonable}. Params denotes the number of model parameters. All reconstruction experiments are per instance INR fitting experiments.

\textbf{Configuration.} SQMM is kept near the QIREN parameter scale and uses five bands with \([2,3,4,4,5]\) reuploads and two data qubits per band. For audio, SQMM encodes the raw coordinate with 16 dyadic sinusoidal frequencies uploaded through \(R_Y\) rotations. For images, SQMM uses the same circuit with 8 consecutive frequencies per axis uploaded through \(R_Y\) and \(R_X\) rotations, and RGB images only widen the output layer. All bands share a two qubit memory register, initialized from the same encoding, with controlled \(R_Y\) reads and \(XY\) writes. Every band is read out through its Pauli \(Z\) and \(ZZ\) expectations, and the decoder is an MLP with one hidden layer of width 5 and GELU activation. Full configurations and training details are given in Appendix~\ref{app:config}.

\subsection{Base Benchmark}\label{exp:Classic_INR}

The classic INR benchmark follows the QIREN protocol. The cello signal is sampled at 1000 points with stride 5, normalized in amplitude, and indexed by timestamps evenly spaced in \([-1,1]\). The image task uses Astronaut, Camera, and Coffee images cropped and downsampled to \(32 \times 32\). This setting tests whether memory coupling improves the small signal regime used in prior quantum INR comparisons. For this benchmark only, models trained for 300 epochs for Cello, Camera and Coffee to match QIREN's benchmark. The column \(\text{SQMM}_{2048}\) estimates every expectation from 2048 measurement shots in training and evaluation (Appendix~\ref{app:finite}). Figure~\ref{fig:cello_wave} shows the Cello reconstruction.

\begin{table*}[t]
\centering
\caption{MSE\(\downarrow\) \((\times 10^{-3})\) on sound and image representation tasks, mean \(\pm\) SD. \(\text{}^\star\) marks our model and its subscript the number of shots. QVF has no audio configuration. Best runs of all models are in Table~\ref{tab:classic-bestrun}. Best in \textbf{bold}.}
\label{tab:classic-inr}
\begingroup
\sqtabsetup
\footnotesize
\begin{tabular*}{\textwidth}{@{\extracolsep{\fill}}lccccccccc@{}}
\toprule
 & $\text{SQMM}^\star$ & $\text{SQMM}^\star_{2048}$ & QIREN & QVF & IGA & FINER & SIREN & MSA-INR & CAFE (enc.) \\
\midrule
Params & 642 & 642 & 649 &  & 673 & 691 & 673 & 673 & 703 \\
Cello & \best{0.48{\scriptsize$\pm$0.08}} & 0.76{\scriptsize$\pm$0.06} & 6.79{\scriptsize$\pm$0.35} &  & 4.87{\scriptsize$\pm$0.01} & 3.73{\scriptsize$\pm$0.44} & 5.02{\scriptsize$\pm$0.08} & 7.04{\scriptsize$\pm$1.23} & 3.60{\scriptsize$\pm$0.60} \\
\midrule
Params & 654 & 654 & 657 & 671 & 687 & 697 & 687 & 697 & 709 \\
Astronaut & \best{1.12{\scriptsize$\pm$0.10}} & 2.86{\scriptsize$\pm$0.53} & 5.90{\scriptsize$\pm$0.49} & 12.61{\scriptsize$\pm$1.65} & 2.01{\scriptsize$\pm$0.53} & 2.80{\scriptsize$\pm$0.31} & 2.15{\scriptsize$\pm$0.22} & 2.03{\scriptsize$\pm$0.28} & 10.46{\scriptsize$\pm$5.92} \\
Camera & \best{0.38{\scriptsize$\pm$0.08}} & 1.27{\scriptsize$\pm$0.16} & 2.19{\scriptsize$\pm$0.31} & 4.43{\scriptsize$\pm$1.38} & 1.07{\scriptsize$\pm$0.46} & 1.96{\scriptsize$\pm$0.62} & 0.76{\scriptsize$\pm$0.47} & 1.14{\scriptsize$\pm$0.14} & 3.36{\scriptsize$\pm$1.41} \\
Coffee & \best{0.78{\scriptsize$\pm$0.05}} & 1.79{\scriptsize$\pm$0.16} & 2.52{\scriptsize$\pm$0.67} & 14.38{\scriptsize$\pm$6.80} & 1.63{\scriptsize$\pm$0.42} & 1.85{\scriptsize$\pm$0.31} & 2.26{\scriptsize$\pm$0.40} & 1.37{\scriptsize$\pm$0.19} & 5.17{\scriptsize$\pm$1.06} \\
\bottomrule
\end{tabular*}
\endgroup
\end{table*}

SQMM obtains the best reconstruction error on each sound and image representation task in Table~\ref{tab:classic-inr}. With 2048 shots SQMM stays best on the audio sample, Cello. The comparison is most informative against QIREN, FINER, and QVF because they represent the closest quantum INR baseline, a strong adaptive classical spectral baseline, and a recent quantum visual field baseline.

\subsection{Model analysis}
\label{sec:model-analysis}

\begin{wraptable}{r}{0.5\linewidth}
  \vspace{-14pt}
\centering
\footnotesize
\setlength{\tabcolsep}{2.5pt}
\caption{Mechanism ablation on Cello, MSE \((\times 10^{-3})\), mean \(\pm\) SD. No memory, local qubits replaces the shared memory with two local qubits in every band and shares nothing between bands. All samples are in Table~\ref{tab:current-winner-memory-spectral-ex}. Best in \textbf{bold}.}
\label{tab:nomem}
\begin{tabular}{lcc}
\toprule
Setting & Params & MSE$\downarrow$ (\(\times10^{-3}\)) \\
\midrule
No memory, local qubits & 650 & 0.53{\scriptsize$\pm$0.06} \\
Memory & 642 & 0.52{\scriptsize$\pm$0.08} \\
$\mathcal L_{\mathrm{out}}$ & 642 & 0.54{\scriptsize$\pm$0.06} \\
$\mathcal L_{\mathrm{align}}$ (Ours) & 642 & \best{0.48{\scriptsize$\pm$0.08}} \\
\bottomrule
\end{tabular}
\vspace{-8pt}

\end{wraptable}
The mechanism ablation isolates the contribution of memory access and spectral alignment. Table~\ref{tab:nomem} compares SQMM on Cello with a no memory control that gives every band two further qubits that act locally and are not shared between bands, which matches the qubit count and parameter budget of SQMM. The no memory control has a higher error than SQMM. On the full classic benchmark (Table~\ref{tab:current-winner-memory-spectral-ex}), adding memory with read write substantially improves reconstruction. Directly supervising the final output spectrum with a control loss \(\mathcal L_{\mathrm{out}}\) (Appendix~\ref{app:loss}) rather than the memory spectrum gives a similar error, while aligning the memory spectrum gives the best result at the same parameter count.

\begin{table*}[!htbp]
\centering
\footnotesize
\setlength{\tabcolsep}{1.5pt}
\caption{SQMM without and with \(\mathcal L_{\mathrm{align}}\) on the base benchmark, mean \(\pm\) SD. Cello reports LSD, the images SSIM and LPIPS. Best in \textbf{bold}.}
\label{tab:base-analysis}
\begin{tabular*}{\textwidth}{@{\extracolsep{\fill}}lcccccccc@{}}
\toprule
 & \multicolumn{2}{c}{Cello} & \multicolumn{2}{c}{Astronaut} & \multicolumn{2}{c}{Camera} & \multicolumn{2}{c}{Coffee} \\
\cmidrule(lr){2-3}\cmidrule(lr){4-5}\cmidrule(lr){6-7}\cmidrule(lr){8-9}
Metric & No loss & \(\mathcal L_{\mathrm{align}}\) & No loss & \(\mathcal L_{\mathrm{align}}\) & No loss & \(\mathcal L_{\mathrm{align}}\) & No loss & \(\mathcal L_{\mathrm{align}}\) \\
\midrule
MSE$\downarrow$ (\(\times10^{-3}\)) & 0.52{\scriptsize$\pm$0.08} & \best{0.48{\scriptsize$\pm$0.08}} & 1.23{\scriptsize$\pm$0.21} & \best{1.12{\scriptsize$\pm$0.10}} & 0.42{\scriptsize$\pm$0.06} & \best{0.38{\scriptsize$\pm$0.08}} & 0.88{\scriptsize$\pm$0.11} & \best{0.78{\scriptsize$\pm$0.05}} \\
PSNR$\uparrow$ (dB) & 32.85{\scriptsize$\pm$0.67} & \best{33.27{\scriptsize$\pm$0.72}} & 35.34{\scriptsize$\pm$0.71} & \best{35.66{\scriptsize$\pm$0.39}} & 39.83{\scriptsize$\pm$0.66} & \best{40.28{\scriptsize$\pm$0.94}} & 36.60{\scriptsize$\pm$0.54} & \best{37.12{\scriptsize$\pm$0.28}} \\
LSD$\downarrow$ (1D) & 4.70{\scriptsize$\pm$0.25} & \best{4.45{\scriptsize$\pm$0.14}} &  &  &  &  &  &  \\
SSIM$\uparrow$ (\(\times10^{-2}\)) (2D) &  &  & 99.23{\scriptsize$\pm$0.10} & \best{99.32{\scriptsize$\pm$0.07}} & 98.21{\scriptsize$\pm$0.26} & \best{98.45{\scriptsize$\pm$0.32}} & 98.93{\scriptsize$\pm$0.10} & \best{99.06{\scriptsize$\pm$0.04}} \\
LPIPS$\downarrow$ (\(\times10^{-4}\)) (2D) &  &  & 3.54{\scriptsize$\pm$0.84} & \best{2.64{\scriptsize$\pm$0.66}} & 3.14{\scriptsize$\pm$1.26} & \best{2.82{\scriptsize$\pm$0.90}} & 3.55{\scriptsize$\pm$0.99} & \best{3.40{\scriptsize$\pm$0.92}} \\
\bottomrule
\end{tabular*}
\end{table*}

Table~\ref{tab:base-analysis} shows the same effect on every sample of the base benchmark. Spectral alignment improves MSE, PSNR, and log spectral distance on Cello, and MSE, PSNR, SSIM, and LPIPS on the images. The consistency across sound and images is more important than the absolute size of any single gain. Additional analysis in Appendix~\ref{app:exp}.

\subsection{Audio}

Audio reconstruction evaluates whether the single signal improvement transfers to varied waveform structure. GTZAN provides music, UrbanSound8K (US8K) provides environmental sounds, and VCTK provides speech. Clips are converted to mono. GTZAN uses 22.05 kHz audio, UrbanSound8K is resampled to 22.05 kHz, and VCTK is resampled to 24 kHz. A stride of 5 keeps 1024 coordinates beginning at 0.5 seconds, using the same coordinate normalization as the cello task.

\begin{table}[t]
\centering
\caption{Audio MSE\(\downarrow\) \((\times 10^{-3})\), mean \(\pm\) SD. \(\text{}^\star\) marks our model. All models are in Table~\ref{tab:audio-full}. Best in \textbf{bold}.}
\label{tab:audio-results}
\begingroup
\sqtabsetup
\footnotesize
\begin{tabular*}{\linewidth}{@{\extracolsep{\fill}}lcccccccc@{}}
\toprule
 & SQMM$^\star$ & QIREN & FINER & SIREN & NeRF & MSA-INR & IGA & CAFE (enc.) \\
\midrule
Params & 642 & 649 & 691 & 673 & 698 & 673 & 673 & 703 \\
\midrule
GTZAN & \best{4.71{\scriptsize$\pm$0.17}} & 44.19{\scriptsize$\pm$1.88} & 5.14{\scriptsize$\pm$1.06} & 8.65{\scriptsize$\pm$0.61} & 8.89{\scriptsize$\pm$0.44} & 42.11{\scriptsize$\pm$3.39} & 8.45{\scriptsize$\pm$0.56} & 6.42{\scriptsize$\pm$1.34} \\
US8K & \best{6.11{\scriptsize$\pm$0.27}} & 51.83{\scriptsize$\pm$0.86} & 8.56{\scriptsize$\pm$2.35} & 15.55{\scriptsize$\pm$1.49} & 13.96{\scriptsize$\pm$0.72} & 52.21{\scriptsize$\pm$1.55} & 15.47{\scriptsize$\pm$1.22} & 10.83{\scriptsize$\pm$3.75} \\
VCTK & \best{1.00{\scriptsize$\pm$0.03}} & 10.70{\scriptsize$\pm$0.10} & 1.66{\scriptsize$\pm$0.21} & 2.98{\scriptsize$\pm$0.15} & 2.59{\scriptsize$\pm$0.15} & 8.51{\scriptsize$\pm$0.90} & 2.69{\scriptsize$\pm$0.13} & 2.25{\scriptsize$\pm$0.34} \\
\bottomrule
\end{tabular*}
\endgroup
\end{table}

Table~\ref{tab:audio-results} shows that SQMM remains stable across the three audio domains. The strongest competing baseline is FINER, which is consistent with the importance of adaptive spectral structure in audio INRs. SQMM improves over FINER on all datasets with fewer parameters.

\subsection{Images}

For image reconstruction, the preprocessing follows Section~\ref{exp:Classic_INR}. We evaluate grayscale CIFAR10 using 100 images. This setting tests whether the SQMM gain extends beyond the three scikit image examples to a larger image collection. To test larger images, SQMM is fitted to the Kodak and DIV2K images from \(32^2\) to \(256^2\) pixels and at full size (Appendix~\ref{app:ladder}).

\begin{figure}[t]
  \centering
  \includegraphics[width=0.74\linewidth]{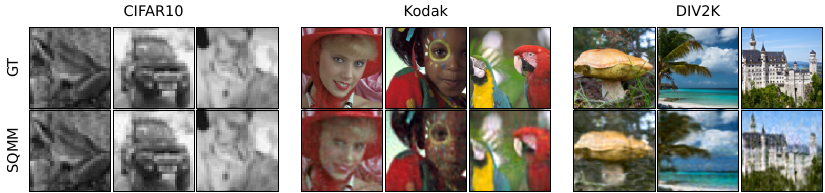}
  \caption{Ground truth (top) and SQMM (bottom), CIFAR10 at \(32\times32\), Kodak and DIV2K at \(128\times128\).}
  \label{fig:main-images}
\end{figure}

\begin{table}[t]
  \centering
\caption{Grayscale CIFAR10, mean \(\pm\) SD. \(\text{}^\star\) marks our model. Table~\ref{tab:cifar-full} has all models. Best in \textbf{bold}.}
\label{tab:cifar10-gray-classical}
\begingroup
\sqtabsetup
\footnotesize
\begin{tabular*}{\linewidth}{@{\extracolsep{\fill}}lccccccc@{}}
\toprule
 & SQMM$^\star$ & QIREN+ & QVF & FINER & SIREN & IGA & MSA-INR \\
\midrule
Params & 654 & 801 & 671 & 697 & 687 & 687 & 697 \\
\midrule
SSIM$\uparrow$ (\(\times10^{-2}\)) & \best{99.21{\scriptsize$\pm$0.03}} & 97.69{\scriptsize$\pm$0.19} & 92.90{\scriptsize$\pm$0.23} & 97.43{\scriptsize$\pm$0.39} & 98.37{\scriptsize$\pm$0.32} & 98.56{\scriptsize$\pm$0.27} & 98.59{\scriptsize$\pm$0.14} \\
LPIPS$\downarrow$ (\(\times10^{-4}\)) & \best{3.44{\scriptsize$\pm$0.25}} & 30.41{\scriptsize$\pm$2.98} & 104.84{\scriptsize$\pm$8.38} & 14.17{\scriptsize$\pm$3.69} & 8.22{\scriptsize$\pm$1.89} & 7.64{\scriptsize$\pm$1.70} & 10.56{\scriptsize$\pm$1.93} \\
MSE$\downarrow$ (\(\times10^{-3}\)) & \best{0.46{\scriptsize$\pm$0.02}} & 1.67{\scriptsize$\pm$0.13} & 5.12{\scriptsize$\pm$0.17} & 1.45{\scriptsize$\pm$0.21} & 0.98{\scriptsize$\pm$0.21} & 0.98{\scriptsize$\pm$0.18} & 0.90{\scriptsize$\pm$0.11} \\
\bottomrule
\end{tabular*}
\endgroup

  \centering
\caption{Kodak and DIV2K, MSE\(\downarrow\) \((\times 10^{-3})\), mean \(\pm\) SD. Kodak without a size is the native \(768\times512\). \(\text{}^\star\) marks our model. All models are in Tables~\ref{tab:ladder-kodak} and~\ref{tab:ladder-div2k}. Best in \textbf{bold}.}
\label{tab:images-main}
\begingroup
\sqtabsetup
\footnotesize
\renewcommand{\arraystretch}{1.0}
\begin{tabular*}{\linewidth}{@{\extracolsep{\fill}}lccccccc@{}}
\toprule
 & SQMM$^\star$ & NeRF & MSA-INR & SIREN & IGA & R+RFF & FLAIR \\
\midrule
Params & 666 & 1023 & 747 & 717 & 717 & 723 & 723 \\
\midrule
Kodak $32^2$ & \best{1.58{\scriptsize$\pm$0.10}} & 3.80{\scriptsize$\pm$0.42} & 1.85{\scriptsize$\pm$0.06} & 2.70{\scriptsize$\pm$0.33} & 3.11{\scriptsize$\pm$0.34} & 3.35{\scriptsize$\pm$0.07} & 3.48{\scriptsize$\pm$0.08} \\
Kodak $64^2$ & \best{5.43{\scriptsize$\pm$0.03}} & 8.76{\scriptsize$\pm$0.68} & 6.84{\scriptsize$\pm$0.06} & 6.89{\scriptsize$\pm$0.26} & 7.06{\scriptsize$\pm$0.25} & 9.47{\scriptsize$\pm$0.18} & 9.57{\scriptsize$\pm$0.46} \\
Kodak $128^2$ & \best{11.54{\scriptsize$\pm$0.05}} & 14.94{\scriptsize$\pm$0.26} & 14.02{\scriptsize$\pm$0.05} & 13.18{\scriptsize$\pm$0.47} & 13.65{\scriptsize$\pm$0.36} & 16.89{\scriptsize$\pm$0.29} & 17.57{\scriptsize$\pm$0.28} \\
Kodak $256^2$ & \best{18.52{\scriptsize$\pm$0.04}} & 21.57{\scriptsize$\pm$0.23} & 21.54{\scriptsize$\pm$0.04} & 20.57{\scriptsize$\pm$0.55} & 20.94{\scriptsize$\pm$0.54} & 24.53{\scriptsize$\pm$0.34} & 25.37{\scriptsize$\pm$0.14} \\
Kodak & \best{32.08{\scriptsize$\pm$0.15}} & 34.79{\scriptsize$\pm$0.40} & 35.64{\scriptsize$\pm$0.26} & 34.44{\scriptsize$\pm$0.61} & 34.04{\scriptsize$\pm$0.56} & 38.54{\scriptsize$\pm$0.64} & 39.04{\scriptsize$\pm$0.39} \\
\midrule
DIV2K $32^2$ & \best{3.13{\scriptsize$\pm$0.08}} & 6.91{\scriptsize$\pm$0.20} & 3.58{\scriptsize$\pm$0.09} & 5.22{\scriptsize$\pm$0.74} & 5.34{\scriptsize$\pm$0.55} & 6.12{\scriptsize$\pm$0.22} & 7.06{\scriptsize$\pm$0.13} \\
DIV2K $64^2$ & \best{11.13{\scriptsize$\pm$0.06}} & 18.57{\scriptsize$\pm$0.66} & 13.66{\scriptsize$\pm$0.11} & 14.37{\scriptsize$\pm$0.75} & 15.22{\scriptsize$\pm$0.60} & 18.42{\scriptsize$\pm$0.41} & 19.83{\scriptsize$\pm$0.28} \\
DIV2K $128^2$ & \best{23.16{\scriptsize$\pm$0.34}} & 30.87{\scriptsize$\pm$0.38} & 27.35{\scriptsize$\pm$0.41} & 27.07{\scriptsize$\pm$1.02} & 28.11{\scriptsize$\pm$0.98} & 32.63{\scriptsize$\pm$0.76} & 35.18{\scriptsize$\pm$0.30} \\
DIV2K $256^2$ & \best{35.15{\scriptsize$\pm$0.43}} & 42.68{\scriptsize$\pm$1.26} & 39.84{\scriptsize$\pm$0.31} & 39.34{\scriptsize$\pm$0.74} & 40.02{\scriptsize$\pm$0.40} & 45.80{\scriptsize$\pm$0.94} & 48.18{\scriptsize$\pm$0.70} \\
DIV2K $512^2$ & \best{54.13{\scriptsize$\pm$0.44}} & 61.08{\scriptsize$\pm$0.53} & 59.10{\scriptsize$\pm$0.46} & 58.41{\scriptsize$\pm$1.14} & 58.40{\scriptsize$\pm$0.39} & 65.14{\scriptsize$\pm$0.85} & 68.11{\scriptsize$\pm$1.31} \\
\bottomrule
\end{tabular*}
\endgroup

\end{table}

Table~\ref{tab:cifar10-gray-classical} shows that SQMM also improves reconstruction on grayscale CIFAR10, where QIREN+ adds one spectrum layer to QIREN. SQMM achieves the lowest MSE and LPIPS and the highest SSIM with the fewest parameters of the compared models. This suggests that the memory coupled representation is not limited to the scikit image setting and transfers to a larger image collection. On Kodak and DIV2K, SQMM obtains the lowest MSE of all models at every resolution, including NeRF which holds \(\times \vNeRFParamRatio\) the parameters of SQMM (Table~\ref{tab:images-main}), and Figure~\ref{fig:main-images} shows example reconstructions.

\section{Discussion}\label{sec:disc}

SQMM suggests that memory coupling is useful not only as an architectural addition, but as a mechanism for organizing spectral computation across quantum INR bands. In prior QINR models, Fourier features are typically produced by independent quantum feature blocks or readout circuits~\citep{zhao2024quantumimplicitneuralrepresentations,Wang2025QVF}. In SQMM, the shared memory provides an intermediate quantum state through which later bands can condition their coefficients on spectra expressed by earlier bands. This connects the architectural coupling of Section~\ref{sec:architecture} to the Fourier view of data reuploading circuits (Section~\ref{sec:theory} and Appendix~\ref{app:proofs})~\citep{schuld2021effect,Casas2023Multidimensional,yu2022power}.

The alignment objective further indicates that the memory should be treated as a trainable spectral interface rather than an opaque hidden state. The ablations in Section~\ref{sec:model-analysis} support this interpretation. Memory access improves reconstruction, and memory alignment gives the strongest result among the compared variants in Table~\ref{tab:current-winner-memory-spectral-ex}.

These results position spectral memory as a practical design principle for compact QINRs, which could extend beyond per instance reconstruction to coordinate based settings with central spectral structure, such as higher dimensional signals, time varying representations, and physical fields~\citep{mildenhall2020nerf,woo2023learning,huang2022universal}. SQMM inherits the practical constraints of parameterized QML and recurrent quantum models, such as noise sensitivity, simulator cost, and hardware dependent scaling with circuit depth and memory size~\citep{mcclean2016theory,Benedetti2019PQCML,Sim2019Expressibility,Bausch2020rqnn,chen2022quantum}. No experiment was run on quantum hardware, and only finite shot estimation and coherent gate perturbations are simulated. At inference the band observables need \(\mathcal O(d_FT/\mathrm{e}^2)\) shots at standard error \(\mathrm{e}\), while full Pauli memory probing needs \(\mathcal O(4^mT(n-1)/\mathrm{e}^2)\) shots per training update (Appendix~\ref{app:lmt}).

\section{Conclusion}\label{sec:con}

SQMM introduces a memory coupled quantum INR where sequential data reuploading bands share a coherent quantum memory. The analysis shows that input independent memory read and write operations preserve the finite Fourier structure of data reuploading circuits while allowing later bands to condition on an operator valued spectral summary of earlier bands.
SQMM achieves the best MSE among the evaluated compact INR and QINR baselines on the considered reconstruction benchmarks. Questions of quantum computational or hardware advantage remain outside the scope of this work. Overall, the results support coherent spectral memory as a useful mechanism for compact quantum implicit representations, with simulation cost and hardware measurement overhead as the main limitations.

\begin{ack}
This work was supported by the D.E. Koshland Jr. Family Career Development Chair in Advanced Technologies in Electrical and Computer Engineering. We declare no competing interests.
\end{ack}

{
\small
\bibliographystyle{plainnat}
\bibliography{references}
}

\newpage

\section{Notation}
\label{app:notation}

\begin{table}[h]
\centering
\small
\setlength{\tabcolsep}{6pt}
\caption{Notation.}
\label{tab:notation}
\begin{tabular}{@{}l>{\raggedright\arraybackslash}p{0.72\linewidth}@{}}
\toprule
Symbol & Meaning \\
\midrule
\(x\in\RR^d\), \(y\), \(\hat y(x)\) & coordinate, target value, prediction \\
\(\PE(x)\), \(\gamma_i(x)\), \(\gamma_M(x)\) & positional encoding, upload variable of band \(i\), one time memory upload \\
\(n\), \(i\), \(L_i\) & number of bands, band index, number of uploads in band \(i\) \\
\(D_i\), \(q_i\), \(M\), \(m\) & local data register of band \(i\) and its qubit count, memory register and its qubit count \\
\(\eta_{D_i}\), \(\sigma_M^{(i)}(x)\) & new data state of band \(i\), memory state after band \(i\) (\(\sigma_M^{(0)}\) is the initial memory) \\
\(\cB_i\), \(\cS_{i,\ell}\), \(\cV_{i,\ell}\) & band channel, input dependent upload layer, input independent (trainable, read, write) block \\
\(\xi_i(x)\) & joint state of \(D_i\otimes M\) after band \(i\) \\
\(O_{i,a}\), \(\bm f_i(x)\), \(d_{f,i}\), \(d_F\) & band observables, band features, their number per band and in total \\
\(G_\theta\) & classical decoder \\
\(\Omega^{\loc}_{i,x}\), \(\Omega^{(i)}_{M,x}\), \(\Omega_{f_i,x}\) & frequency support of band \(i\), of the memory after band \(i\), of the features of band \(i\) \\
\(\tau^{(i)}_\mu\) & matrix valued Fourier coefficient of \(\sigma_M^{(i)}\) at frequency \(\mu\) \\
\(\cB_{i,\nu}\), \(\ell_{i,a,\nu}\) & Fourier coefficient of the band channel, linear readout functional at frequency \(\nu\) \\
\(c^{(i)}_{\omega,a}\) & Fourier coefficient of feature \(a\) of band \(i\) \\
\(\Omega_{\mathrm{grid}}=\{\bar\omega_h\}_{h=1}^{N_\omega}\) & finite frequency grid of the alignment loss \\
\(\widehat a(u)\), \(\Phi\), \(\varepsilon\) & estimated Fourier coefficients of a sampled signal \(u\), design matrix, ridge (\(\varepsilon=0\) in all experiments) \\
\(\mathcal Q_m\), \(Q_s\), \(\chi^M_{i,s}\) & \(m\) qubit Pauli strings, one string, memory Pauli signal \(\Tr[Q_s\sigma_M^{(i)}]\) \\
\(P^F_i\), \(P^M_i\) & accumulated prefix feature spectrum, memory occupancy spectrum at boundary \(i\) \\
\(\mathcal N_\delta\), \(\delta\) & normalization to unit mass with smoothing \(\delta\) \\
\(\lambda_{\mathrm{align}}\), \(\lambda_{\mathrm{out}}\) & weights of \(\mathcal L_{\mathrm{align}}\) and of the output spectral control \(\mathcal L_{\mathrm{out}}\) \\
\(\operatorname{sg}\) & stop gradient \\
\bottomrule
\end{tabular}
\end{table}

\section{Quantum and Fourier Preliminaries}
\label{app:prelim}

This appendix provides additional background to Section~\ref{sec:background} and the definitions and proofs used by Section~\ref{sec:method}.
The basic quantum notation is included to make the method readable to non QML reviewers.

\paragraph{Qubits, registers, and density matrices.}
A qubit is a two level quantum system with computational basis states \(\ket{0}\) and \(\ket{1}\).
A pure one qubit state has the form \(\ket{\psi}=\alpha\ket{0}+\beta\ket{1}\), with \(|\alpha|^2+|\beta|^2=1\).
An \(n_q\) qubit register lives in \((\mathbb C^2)^{\otimes n_q}\), whose dimension is \(2^{n_q}\).
When a subsystem may be entangled with another subsystem, it is described by a density matrix \(\rho\), a positive semidefinite operator with \(\Tr[\rho]=1\).
A register is a group of qubits. If \(\rho_{DM}\) is a joint state on a data register \(D\) with \(q\) qubits and memory register \(M\) with \(m\) qubits, then \(\rho_{DM}\in\mathbb C^{2^{q+m}\times 2^{q+m}}\), and the reduced memory state is \(\sigma_M=\Tr_D[\rho_{DM}]\in\mathbb C^{2^m\times 2^m}\).

\paragraph{Gates and channels.}
Quantum gates are unitary transformations.
Typical single qubit rotation gates are
\begin{equation}
R_X(\theta)=e^{-\mathrm{i}\theta X/2},
\qquad
R_Y(\theta)=e^{-\mathrm{i}\theta Y/2},
\qquad
R_Z(\theta)=e^{-\mathrm{i}\theta Z/2},
\end{equation}
where \(X,Y,Z\in\mathbb C^{2\times2}\) are Pauli matrices.
Two qubit gates, such as CNOT, CZ, or trainable Ising couplings, create correlations between qubits.
For the proofs, it is convenient to treat a circuit block as a linear channel \(\mathcal C\) acting on density matrices.
Unitary gates are the special case \(\mathcal C(\rho)=U\rho U^\dagger\).

\paragraph{Partial trace identities.}
For registers \(D\) and \(M\), the partial trace \(\Tr_D\) is the unique linear map satisfying
\[
\Tr\!\left[A_M\,\Tr_D(X_{DM})\right]
=
\Tr\!\left[(I_D\otimes A_M)X_{DM}\right]
\]
for every operator \(A_M\) on \(M\).
If \(K\) acts only on \(D\), then
\[
\Tr_D\!\left[(K\otimes I_M)X_{DM}\right]
=
\Tr_D\!\left[X_{DM}(K\otimes I_M)\right].
\]
This is the cyclicity of trace inside the subsystem being traced.
It is used in Proposition~\ref{prop:memory-readout}.

\paragraph{Measurement and expectation values.}
For a Hermitian observable \(O\) and state \(\rho\), the expectation value is \(\langle O\rangle=\Tr[O\rho]\).
SQMM uses expectation valued Pauli readouts as features.
These features can be estimated either exactly in statevector simulation or by repeated circuit executions on hardware.
They should not be interpreted as single shot selective measurement outcomes from the same coherent memory trajectory.
The formal nonselective measurement statement is given in Proposition~\ref{prop:memory-readout}.

\paragraph{Kraus form of local measurements.}
A nonselective measurement on a data register \(D\) is represented by Kraus operators \(\{K_\zeta\}_\zeta\) on \(D\) that satisfy
\[
\sum_\zeta K_\zeta^\dagger K_\zeta=I_D.
\]
The post measurement joint state averaged over outcomes is
\[
\sum_\zeta (K_\zeta\otimes I_M)\rho_{DM}(K_\zeta^\dagger\otimes I_M).
\]
When one conditions on a particular outcome, the corresponding branch must be divided by its probability.
The distinction between averaged and conditioned states is the reason SQMM treats readouts as expectation valued probes rather than as reused single shot outcomes.

\paragraph{Operator spaces and channel coefficients.}
Let \(\mathcal L(\mathcal H)\) denote the complex vector space of linear operators on a finite dimensional Hilbert space \(\mathcal H\).
A quantum channel is a completely positive trace preserving linear map on \(\mathcal L(\mathcal H)\).
Although a channel maps density matrices to density matrices, its linear extension acts on arbitrary operators in \(\mathcal L(\mathcal H)\).
This convention is used throughout the proofs when Fourier coefficient maps such as \(\cB_{i,\nu}\) act on matrices \(\eta_{D_i}\otimes\tau_\mu^{(i)}\).
The coefficient maps in a Fourier expansion of a channel need not themselves be physical channels.
They are linear superoperators whose phase weighted sum is the physical channel at each coordinate value.

\paragraph{Data reuploading and coordinate Fourier support.}
Data reuploading circuits repeatedly inject the uploaded variable \(\gamma_i(x)\in\RR^{g_i}\) into parameterized gates, interleaved with trainable input independent layers. For formulas in the upload domain, write \(z\in\RR^{g_i}\) for a generic uploaded variable and reserve \(\gamma_i(x)\) for the coordinate dependent value. Each input dependent upload gate is generated by a finite dimensional Hermitian operator. A typical component gate has the form
\begin{equation}
U_{i,\ell,r}(z)
=
\exp\left(-\mathrm{i}z_r H_{i,\ell,r}\right),
\qquad
r=1,\ldots,g_i,
\end{equation}
where \(H_{i,\ell,r}\) is the encoding generator for upload component \(r\). Since \(H_{i,\ell,r}\) has a finite spectrum, the matrix entries of this gate are finite sums of phases whose frequencies are the negated eigenvalues of \(H_{i,\ell,r}\), so the induced channel and its expectation values have frequencies given by eigenvalue gaps. Across layers, accessible upload frequencies are finite sums of such gaps \citep{schuld2021effect,Casas2023Multidimensional,yu2022power}.

\paragraph{Linear algebra notation.}
For a complex matrix \(A\), \(A^\dagger\) denotes the conjugate transpose.
The Frobenius norm is \(\|A\|_F^2=\Tr(A^\dagger A)\).
The Hilbert Schmidt inner product is
\[
\langle A,B\rangle_{\mathrm{HS}}=\Tr(A^\dagger B).
\]
The ridge estimate in Eq.~\eqref{eq:fourier-ls-app} is computed on the chosen finite frequency grid only.
When \(\varepsilon=0\), exact coefficient recovery on that grid requires \(\Phi\) to have full column rank.
When \(\varepsilon>0\), the estimator is a stabilized projection and its coefficients are biased toward zero.

\paragraph{Fourier supports and conjugation.}
A finite operator valued Fourier expansion over coordinates has the form
\[
F(x)=\sum_{\omega\in\Omega} e^{\mathrm{i}\langle\omega,x\rangle}F_\omega,
\]
where \(\Omega\subset\RR^d\) is finite and each \(F_\omega\) is independent of \(x\).
The set \(\Omega\) may be a superset of the nonzero coefficient support.
Consequently, all support statements in the proofs are upper bounds.
If \(F(x)\) has support \(\Omega\), then \(F(x)^\dagger\) has support contained in \(-\Omega\), with coefficient \(F_\omega^\dagger\) at frequency \(-\omega\).
Thus products such as \(U(x)\eta U(x)^\dagger\) remain finite Fourier whenever \(U(x)\) is finite Fourier.

In the uploaded variable domain one may write
\begin{equation}
\cS_{i,\ell}(z)
=
\sum_{\kappa\in\Omega_{i,\ell,\gamma}}
e^{\mathrm{i}\langle \kappa,z\rangle}
\cS_{i,\ell,\kappa},
\qquad
\Omega_{i,\ell,\gamma}\subset\RR^{g_i}.
\end{equation}
After substituting \(z=\gamma_i(x)=A_i\PE(x)+b_i\), the layer is analyzed on the coordinate domain through
\begin{equation}
\cS_{i,\ell}(\gamma_i(x))
=
\sum_{\nu\in\Omega_{i,\ell,x}}
e^{\mathrm{i}\langle \nu,x\rangle}
\cS_{i,\ell,\nu}.
\end{equation}
For affine raw coordinate uploads this expansion is exact, with coordinate frequencies induced by \(A_i^\top\kappa\). For fixed sinusoidal positional encodings, the analysis can instead be viewed in the lifted coordinate \(\PE(x)\), or projected onto a finite coordinate grid. If an implementation places a trigonometric polynomial inside a rotation angle, the resulting coordinate expansion may be countable rather than finite. In that case the same support algebra applies to any finite truncation, and the alignment loss uses exactly such a finite working grid.

\paragraph{Pauli basis normalization.}
The \(m\) qubit Pauli strings form an orthogonal basis of \(\mathcal L((\mathbb C^2)^{\otimes m})\) under the Hilbert Schmidt inner product.
With the unnormalized convention used here,
\[
\Tr(Q_sQ_{s'})=2^m\mathbf 1\{s=s'\}.
\]
Hence every operator \(A\) on \(m\) qubits admits the expansion
\[
A=2^{-m}\sum_{Q_s\in\mathcal Q_m}\Tr(Q_sA)Q_s.
\]
Parseval identity for this orthogonal basis gives
\[
\|A\|_F^2
=
2^{-m}\sum_{Q_s\in\mathcal Q_m}|\Tr(Q_sA)|^2.
\]

\section{Proofs for Spectral Memory Propagation}
\label{app:proofs}

This section proves the statements summarized in Section~\ref{sec:theory}.
The proofs use the coordinate domain notation of the method.
Thus \(x\in\RR^d\), \(\gamma_i(x)\in\RR^{g_i}\), coordinate frequencies lie in \(\RR^d\), and phases have the form \(e^{\mathrm{i}\langle\omega,x\rangle}\).
All local supports \(\Omega_{i,x}^{\loc}\) are effective coordinate supports after composing the uploaded variable map \(x\mapsto\gamma_i(x)\).
Unbarred frequencies \(\mu,\nu,\omega\) denote theoretical coordinate frequencies. Barred frequencies \(\bar\omega_h\) are reserved for the finite grid used in Appendix~\ref{app:align-details}. The symbol \(\tau_\omega^{(i)}\) always denotes the matrix Fourier coefficient of the memory density matrix at frequency \(\omega\).
Throughout, \(D_i\) denotes the data register of band~\(i\), \(M\) the shared memory register, \(I_M\) the identity on memory, \(\eta_{D_i}\) the fixed new data state, \(\cS_{i,\ell}(\gamma_i(x))\) the input dependent encoding layers, and \(\cV_{i,\ell}\) the input independent trainable layers.

\begin{lemma}[closure under composition]
\label{lem:closure}
Let \(\mathcal A(x)\) and \(\mathcal C(x)\) be channel families with finite coordinate domain Fourier expansions
\begin{equation}
\mathcal A(x)=\sum_{\alpha\in\Omega_{\mathcal A}} e^{\mathrm{i}\langle\alpha,x\rangle}\mathcal A_\alpha,
\qquad
\mathcal C(x)=\sum_{\beta\in\Omega_{\mathcal C}} e^{\mathrm{i}\langle\beta,x\rangle}\mathcal C_\beta,
\end{equation}
where \(\Omega_{\mathcal A},\Omega_{\mathcal C}\subset\RR^d\) are finite and each coefficient map is independent of \(x\).
Then
\begin{equation}
(\mathcal A\circ \mathcal C)(x)
=
\sum_{\omega\in\Omega_{\mathcal A}+\Omega_{\mathcal C}} e^{\mathrm{i}\langle\omega,x\rangle}
\left(
\sum_{\alpha+\beta=\omega}\mathcal A_\alpha\circ\mathcal C_\beta
\right).
\end{equation}
In particular, the composition is finite Fourier on the coordinate domain.
\end{lemma}

\begin{proof}
Expanding the composition gives
\begin{equation}
(\mathcal A\circ \mathcal C)(x)
=
\sum_{\alpha\in\Omega_{\mathcal A}}\sum_{\beta\in\Omega_{\mathcal C}}
e^{\mathrm{i}\langle\alpha+\beta,x\rangle}
\mathcal A_\alpha\circ\mathcal C_\beta.
\end{equation}
Grouping terms with the same total frequency \(\omega=\alpha+\beta\) gives the stated expression.
\end{proof}

\begin{proposition}[band closure under memory read and write insertion]
\label{prop:band-closure}
Consider a bare data reuploading band
\begin{equation}
\cB_i(\gamma_i(x))
=
\cW_{i,L_i+1}\circ \cS_{i,L_i}(\gamma_i(x))\circ \cW_{i,L_i}\circ\cdots\circ \cS_{i,1}(\gamma_i(x))\circ \cW_{i,1},
\end{equation}
where each \(\cS_{i,\ell}(\gamma_i(x))\) is input dependent and each \(\cW_{i,\ell}\) is input independent.
After inserting any input independent memory read and memory write layers, the resulting band can still be written as
\begin{equation}
\cB_i(\gamma_i(x))
=
\cV_{i,L_i+1}\circ \cS_{i,L_i}(\gamma_i(x))\circ \cV_{i,L_i}\circ\cdots\circ \cS_{i,1}(\gamma_i(x))\circ \cV_{i,1},
\end{equation}
for suitable input independent blocks \(\cV_{i,\ell}\).
Consequently, memory coupling preserves the reuploading form on the enlarged register \(D_i\otimes M\).
\end{proposition}

\begin{proof}
Each inserted read or write layer is independent of \(x\) and \(\gamma_i(x)\), so it can be absorbed into the nearest neighboring input independent block.
For example, a read layer \(\mathcal R\) placed after \(\cS_{i,\ell}(\gamma_i(x))\) can be absorbed by redefining \(\cV_{i,\ell+1}=\cW_{i,\ell+1}\circ\mathcal R\).
A write layer \(\mathcal W\) placed after the last encoding layer is absorbed into the closing block \(\cV_{i,L_i+1}\). The same argument applies to memory only layers.
The only input dependent objects remain the encoding channels \(\cS_{i,\ell}(\gamma_i(x))\), so the alternating reuploading structure is unchanged.
\end{proof}

\begin{corollary}[finite Fourier stack]
\label{cor:stack-induction}
If every band \(\cB_j(\gamma_j(x))\) has a finite effective coordinate Fourier expansion with support \(\Omega_{j,x}^{\loc}\), then every finite prefix channel through the first \(i\) bands has a finite effective coordinate Fourier expansion.
\end{corollary}

\begin{proof}
The case \(i=1\) is immediate.
The inductive step follows from Lemma~\ref{lem:closure} applied to the next band and the existing prefix channel.
Input independent initialization, partial trace operations, and tensoring with a fixed new data state are linear maps independent of \(x\), so they do not introduce new coordinate frequencies.
An encoded initialization has a finite expansion by Proposition~\ref{prop:memory-init-encoding}.
\end{proof}

\begin{proposition}[memory initialization with optional initial encoding]
\label{prop:memory-init-encoding}
Let the memory register have \(m\) qubits.
If the initial memory state is \(\eta_M=\ket{0}\!\bra{0}^{\otimes m}\) or \(\eta_M=\ket{+}\!\bra{+}^{\otimes m}\), and no input dependent gate is applied to memory before the first band, then
\begin{equation}
\sigma_M^{(0)}(x)=\eta_M,
\qquad
\Omega_{M,x}^{(0)}=\{0\}.
\end{equation}
If an input dependent memory encoding with the positional encoding of the bands is applied once at the beginning, define \(\gamma_M(x)=A_M\PE(x)+b_M\), let \(U_M(\gamma_M(x))\) be a finite product of memory upload gates, for example gates \(R_Y((\gamma_M(x))_k)\) on memory qubits, and write \(U_M(x):=U_M(\gamma_M(x))\). Then
\begin{equation}
\sigma_M^{(0)}(x)
=
U_M(x)\eta_M U_M(x)^\dagger
=
\sum_{\mu\in\Omega_{M,x}^{(0)}}
e^{\mathrm{i}\langle\mu,x\rangle}\tau_\mu^{(0)}
\end{equation}
for some finite support \(\Omega_{M,x}^{(0)}\subset\RR^d\) and matrices \(\tau_\mu^{(0)}\in\mathbb C^{2^m\times2^m}\), or approximately for a truncated support.
The memory recursion and the downstream readout proof remain valid with this nontrivial initial support.
\end{proposition}

\begin{proof}
For \(\eta_M=\ket{0}\!\bra{0}^{\otimes m}\) or \(\eta_M=\ket{+}\!\bra{+}^{\otimes m}\) without input dependent preprocessing, \(\sigma_M^{(0)}(x)\) is constant in \(x\).
A constant function has the Fourier expansion \(e^{\mathrm{i}\langle0,x\rangle}\eta_M\), so its support is \(\{0\}\).

Now suppose that the memory is passed through an input dependent unitary \(U_M(\gamma_M(x))\) once at the beginning, with \(\gamma_M(x)=A_M\PE(x)+b_M\).
For a single gate \(R_Y((\gamma_M(x))_k)\),
\begin{equation}
R_Y((\gamma_M(x))_k)
=
\cos\!\left(\frac{(\gamma_M(x))_k}{2}\right)I
-\mathrm{i}\sin\!\left(\frac{(\gamma_M(x))_k}{2}\right)Y.
\end{equation}
Under the same finite or truncated coordinate support assumption used for the data uploading gates, each matrix entry of \(R_Y((\gamma_M(x))_k)\) has a finite or truncated Fourier expansion over \(x\).
Products of finitely many such memory encoding gates also have finite or truncated expansions by the argument of Lemma~\ref{lem:closure}.
The density matrix \(U_M(x)\eta_M U_M(x)^\dagger\) is formed by products of entries from \(U_M(x)\) and their complex conjugates.
Therefore it has a finite or truncated operator valued Fourier expansion with some support \(\Omega_{M,x}^{(0)}\).

The proof of Proposition~\ref{prop:memory-recursion} only assumes that the incoming memory has an operator valued Fourier expansion.
It does not require the base support to be \(\{0\}\).
Therefore the same recursion applies when the memory is initialized by \(\ket{0}\) or \(\ket{+}\) and then receives an initial positional encoding driven by the same \(\PE(x)\) used by the bands.
The only change is that the global support starts from \(\Omega_{M,x}^{(0)}\) rather than \(\{0\}\), so the theory is unchanged after replacing the trivial base support by this encoded base support.
\end{proof}

\begin{proposition}[operator valued Fourier recursion on memory]
\label{prop:memory-recursion}
Assume the memory entering band~\(i+1\) has the form
\begin{equation}
\sigma_M^{(i)}(x)=
\sum_{\mu\in\Omega_{M,x}^{(i)}}
e^{\mathrm{i}\langle\mu,x\rangle}\tau_\mu^{(i)},
\qquad
\Omega_{M,x}^{(i)}\subset\RR^d,
\qquad
\tau_\mu^{(i)}\in\mathbb C^{2^m\times2^m},
\end{equation}
and that band~\(i+1\) has the effective coordinate domain channel expansion
\begin{equation}
\cB_{i+1}(\gamma_{i+1}(x))
=
\sum_{\nu\in\Omega_{i+1,x}^{\loc}}
e^{\mathrm{i}\langle\nu,x\rangle}\cB_{i+1,\nu},
\qquad
\Omega_{i+1,x}^{\loc}\subset\RR^d.
\end{equation}
Then the outgoing memory state
\begin{equation}
\sigma_M^{(i+1)}(x)
=
\Tr_{D_{i+1}}\!\left[
\cB_{i+1}(\gamma_{i+1}(x))
\bigl(\eta_{D_{i+1}}\otimes\sigma_M^{(i)}(x)\bigr)
\right]
\end{equation}
has the expansion
\begin{equation}
\sigma_M^{(i+1)}(x)
=
\sum_{\omega\in\Omega_{M,x}^{(i+1)}}
e^{\mathrm{i}\langle\omega,x\rangle}\tau_\omega^{(i+1)},
\end{equation}
where
\begin{equation}
\tau_\omega^{(i+1)}
=
\sum_{\mu+\nu=\omega}
\Tr_{D_{i+1}}\!\left[
\cB_{i+1,\nu}
\bigl(\eta_{D_{i+1}}\otimes\tau_\mu^{(i)}\bigr)
\right].
\label{eq:memory-recursion-app}
\end{equation}
Consequently,
\begin{equation}
\Omega_{M,x}^{(i+1)}
\subseteq
\Omega_{M,x}^{(i)}+
\Omega_{i+1,x}^{\loc}.
\end{equation}
\end{proposition}

\begin{proof}
Substituting the two Fourier expansions into the memory update gives
\begin{align}
\sigma_M^{(i+1)}(x)
&=
\Tr_{D_{i+1}}\!\left[
\left(\sum_\nu e^{\mathrm{i}\langle\nu,x\rangle}\cB_{i+1,\nu}\right)
\left(\eta_{D_{i+1}}\otimes
\sum_\mu e^{\mathrm{i}\langle\mu,x\rangle}\tau_\mu^{(i)}\right)
\right]
\\
&=
\sum_{\mu,\nu}e^{\mathrm{i}\langle\mu+\nu,x\rangle}
\Tr_{D_{i+1}}\!\left[
\cB_{i+1,\nu}
\bigl(\eta_{D_{i+1}}\otimes\tau_\mu^{(i)}\bigr)
\right].
\end{align}
Regrouping terms by \(\omega=\mu+\nu\) gives Eq.~\eqref{eq:memory-recursion-app} and the support bound.
\end{proof}

\begin{proposition}[downstream readout coefficients]
\label{prop:downstream-coeffs}
Under the assumptions of Proposition~\ref{prop:memory-recursion}, let \(O_{i+1,a}\) be a Hermitian observable used for feature component \(a\) of band \(i+1\).
The feature extraction map
\begin{equation}
f_{i+1,a}(x)
=
\Tr\!\left[
(O_{i+1,a}\otimes I_M)\,
\cB_{i+1}(\gamma_{i+1}(x))
\bigl(\eta_{D_{i+1}}\otimes\sigma_M^{(i)}(x)\bigr)
\right]
\end{equation}
admits a finite or truncated Fourier expansion whose support is contained in
\(\Omega_{M,x}^{(i)}+\Omega_{i+1,x}^{\loc}\). Namely,
\begin{equation}
f_{i+1,a}(x)
=
\sum_{\omega\in\Omega_{M,x}^{(i)}+
\Omega_{i+1,x}^{\loc}}
c_{\omega,a}^{(i+1)}e^{\mathrm{i}\langle\omega,x\rangle},
\end{equation}
with coefficients
\begin{equation}
c_{\omega,a}^{(i+1)}
=
\sum_{\mu+\nu=\omega}\ell_{i+1,a,\nu}(\tau_\mu^{(i)}),
\qquad
\ell_{i+1,a,\nu}(X)
=
\Tr\!\left[
(O_{i+1,a}\otimes I_M)\,
\cB_{i+1,\nu}
\bigl(\eta_{D_{i+1}}\otimes X\bigr)
\right].
\label{eq:coeff-recursion-app}
\end{equation}
Here \(\ell_{i+1,a,\nu}\) is the readout functional associated with observable \(O_{i+1,a}\) and band frequency coefficient \(\cB_{i+1,\nu}\). It takes a memory side matrix \(X\), appends the fixed data state, applies the channel coefficient, and returns a scalar expectation value.
\end{proposition}

\begin{proof}
The proof repeats the expansion in Proposition~\ref{prop:memory-recursion}, but applies the feature extraction functional rather than tracing out the data register. The functional is linear because tensoring with a fixed operator, applying \(\cB_{i+1,\nu}\), and taking a trace are linear in \(X\). Linearity gives Eq.~\eqref{eq:coeff-recursion-app}.
\end{proof}

\begin{corollary}[global support bound]
\label{cor:global-support}
If every band satisfies the assumptions of Proposition~\ref{prop:memory-recursion} and the memory is initialized independently of \(x\), so \(\Omega_{M,x}^{(0)}=\{0\}\), then
\begin{equation}
\Omega_{M,x}^{(i)}
\subseteq
\sum_{j=1}^{i}\Omega_{j,x}^{\loc},
\qquad
\Omega_{f_{i+1},x}
\subseteq
\sum_{j=1}^{i+1}\Omega_{j,x}^{\loc}.
\end{equation}
If the memory receives an initial input dependent encoding as in Proposition~\ref{prop:memory-init-encoding}, then
\begin{equation}
\Omega_{M,x}^{(i)}
\subseteq
\Omega_{M,x}^{(0)}+
\sum_{j=1}^{i}\Omega_{j,x}^{\loc},
\qquad
\Omega_{f_{i+1},x}
\subseteq
\Omega_{M,x}^{(0)}+
\sum_{j=1}^{i+1}\Omega_{j,x}^{\loc}.
\end{equation}
\end{corollary}

\begin{proof}
Apply Proposition~\ref{prop:memory-recursion} recursively.
For input independent initialization, Proposition~\ref{prop:memory-init-encoding} gives \(\Omega_{M,x}^{(0)}=\{0\}\).
For an initial memory encoding driven by the same \(\PE(x)\) used by the bands, Proposition~\ref{prop:memory-init-encoding} gives the finite or truncated base support \(\Omega_{M,x}^{(0)}\).
Applying Proposition~\ref{prop:downstream-coeffs} to the next readout gives the output support bound.
\end{proof}

\section{Readout Probes and Memory Semantics}
\label{app:readout}

The main text separates coherent memory propagation from expectation valued readout.
This section states the formal measurement fact behind that separation.

\begin{proposition}[memory state after nonselective local readout]
\label{prop:memory-readout}
Let \(\rho_{DM}\in\mathbb C^{2^{q+m}\times2^{q+m}}\) be the joint post write state of a data register \(D\) with \(q\) qubits and memory register \(M\) with \(m\) qubits, and let \(\sigma_M=\Tr_D[\rho_{DM}]\).
Let a nonselective measurement act only on \(D\), with Kraus operators \(\{K_\zeta\}_\zeta\) satisfying \(\sum_\zeta K_\zeta^\dagger K_\zeta=I_D\).
Then the reduced memory state after averaging over outcomes is unchanged.
\begin{equation}
\sum_\zeta\Tr_D\!\left[(K_\zeta\otimes I_M)\rho_{DM}(K_\zeta^\dagger\otimes I_M)\right]
=
\sigma_M.
\end{equation}
\end{proposition}

\begin{proof}
Using cyclicity inside the traced subsystem,
\begin{align}
&\sum_\zeta\Tr_D\!\left[(K_\zeta\otimes I_M)\rho_{DM}(K_\zeta^\dagger\otimes I_M)\right]
\\
&\qquad=
\sum_\zeta\Tr_D\!\left[\rho_{DM}(K_\zeta^\dagger K_\zeta\otimes I_M)\right]
=
\Tr_D\!\left[\rho_{DM}\left(\sum_\zeta K_\zeta^\dagger K_\zeta\otimes I_M\right)\right]
=
\Tr_D[\rho_{DM}].
\end{align}
\end{proof}

\paragraph{Selective outcomes.}
If one conditions on a particular outcome \(\zeta\), the memory state becomes
\begin{equation}
\sigma_{M|\zeta}
=
\frac{\Tr_D[(K_\zeta\otimes I_M)\rho_{DM}(K_\zeta^\dagger\otimes I_M)]}{p_\zeta},
\qquad
p_\zeta=\Tr[(K_\zeta^\dagger K_\zeta\otimes I_M)\rho_{DM}],
\end{equation}
which generally depends on \(\zeta\).
Therefore, SQMM features are expectation values estimated from readout probes or separate repeated executions, not single shot selective outcomes whose post measurement memory state is reused downstream.

\section{Details of Spectral Memory Alignment}
\label{app:align-details}

This section expands the vector notation used in Section~\ref{sec:loss}. The propagation lemma uses unbarred theoretical frequencies, while the loss is evaluated only on the finite grid \(\Omega_{\mathrm{grid}}=\{\bar\omega_1,\ldots,\bar\omega_{N_\omega}\}\subset\RR^d\).

Let \(x_1,\ldots,x_T\in\RR^d\) be the coordinate grid used only for the auxiliary spectral loss. For a sampled scalar signal \(u=(u(x_1),\ldots,u(x_T))^\top\in\mathbb C^T\), define the Fourier design matrix \(\Phi\in\mathbb C^{T\times N_\omega}\) by
\begin{equation}
\Phi_{th}=e^{\mathrm{i}\langle\bar\omega_h,x_t\rangle}.
\end{equation}
Coefficients on \(\Omega_{\mathrm{grid}}\) are estimated by ridge stabilized least squares,
\begin{equation}
\widehat a(u)=(\Phi^\dagger\Phi+\varepsilon I_{N_\omega})^{-1}\Phi^\dagger u.
\label{eq:fourier-ls-app}
\end{equation}
Here \(I_{N_\omega}\in\RR^{N_\omega\times N_\omega}\) is the identity matrix. In exact simulation, \(\varepsilon\) can be set to zero when \(\Phi\) has full column rank. For finite shot estimates or ill conditioned grids, \(\varepsilon>0\) stabilizes the coefficient estimate.

For feature component \(a\) of band \(j\), define the sampled signal
\begin{equation}
u_{j,a}^F=(f_{j,a}(x_1),\ldots,f_{j,a}(x_T))^\top,
\qquad
\widehat c^F_{j,a,h}=\bigl[\widehat a(u_{j,a}^F)\bigr]_h.
\end{equation}
Collecting all feature components of band \(j\) at grid frequency \(\bar\omega_h\) gives
\begin{equation}
\widehat{\bm c}_{j,h}^F=
(\widehat c^F_{j,1,h},\ldots,\widehat c^F_{j,d_{f,j},h})
\in\mathbb C^{d_{f,j}}.
\end{equation}
The prefix feature occupancy used in the main text is therefore
\begin{equation}
P_i^F(\bar\omega_h)=
\sum_{j=1}^{i}\|\widehat{\bm c}_{j,h}^F\|_2^2
=
\sum_{j=1}^{i}\sum_{a=1}^{d_{f,j}}|\widehat c^F_{j,a,h}|^2.
\end{equation}
This is an occupancy measure over extracted quantum features, not a spectrum of a linear layer or deployed decoder contribution. The occupancy target records which frequencies have been represented by earlier band features before the final decoder.

To estimate the memory spectrum, use the unnormalized Pauli string basis
\begin{equation}
\mathcal Q_m=\{I,X,Y,Z\}^{\otimes m},
\qquad
|\mathcal Q_m|=4^m,
\qquad
\Tr(Q_sQ_{s'})
=
2^m\mathbf 1\{s=s'\}.
\end{equation}
For each basis element, define the sampled memory signal
\begin{equation}
u_{i,s}^M
=
(\Tr[Q_s\sigma_M^{(i)}(x_1)],\ldots,\Tr[Q_s\sigma_M^{(i)}(x_T)])^\top,
\qquad
\widehat c^M_{i,s,h}
=
\bigl[\widehat a(u_{i,s}^M)\bigr]_h.
\end{equation}
Collecting the Pauli probe coefficients at grid frequency \(\bar\omega_h\) gives
\begin{equation}
\widehat{\bm c}_{i,h}^M
=
(\widehat c^M_{i,s,h})_{Q_s\in\mathcal Q_m}
\in\mathbb C^{4^m}.
\end{equation}
The memory occupancy spectrum is
\begin{equation}
P_i^M(\bar\omega_h)
=
2^{-m}
\sum_{Q_s\in\mathcal Q_m}
|\widehat c^M_{i,s,h}|^2.
\label{eq:memory-spectrum-loss-app}
\end{equation}

If the memory state has the exact expansion
\begin{equation}
\sigma_M^{(i)}(x)
=
\sum_{\omega\in\Omega_{M,x}^{(i)}}
e^{\mathrm{i}\langle\omega,x\rangle}
\tau_\omega^{(i)},
\end{equation}
then, when every frequency of \(\Omega_{M,x}^{(i)}\) is a discrete Fourier frequency of the sampling grid and \(\varepsilon=0\),
\begin{equation}
\widehat c^M_{i,s,h}
=
\Tr[Q_s\tau_{\bar\omega_h}^{(i)}].
\end{equation}
The Pauli strings form an orthogonal operator basis under the Hilbert Schmidt inner product. For any operator \(A\) on \(m\) qubits,
\begin{equation}
A
=
2^{-m}
\sum_{Q_s\in\mathcal Q_m}
\Tr(Q_sA)Q_s,
\end{equation}
and therefore
\begin{equation}
\|A\|_F^2
=
\Tr(A^\dagger A)
=
2^{-m}
\sum_{Q_s\in\mathcal Q_m}
|\Tr(Q_sA)|^2.
\end{equation}
Applying this identity to \(A=\tau_{\bar\omega_h}^{(i)}\) gives
\begin{equation}
P_i^M(\bar\omega_h)
=
\|\tau_{\bar\omega_h}^{(i)}\|_F^2
\end{equation}
up to finite grid estimation error. The identity Pauli string contributes only to the zero frequency trace component in the ideal normalized state setting.

\paragraph{Approximate and truncated spectra.}
The exact identity above requires an exact finite expansion on the coordinate domain and a frequency grid containing the frequency of interest. If the composed circuit has countable or continuous spectral content, or if relevant frequencies are absent from \(\Omega_{\mathrm{grid}}\), the estimator computes the ridge projection of the sampled memory signals onto the chosen grid. In that case \(P_i^M\) is a proxy for the resolved memory spectrum rather than the full memory spectrum. The resulting limitations are standard finite grid spectral limitations, such as, off grid frequencies can leak into nearby grid coefficients, unresolved high frequencies are ignored, insufficient or nonuniform sampling can cause aliasing, and the ridge parameter can bias small coefficients. The propagation lemma still describes the algebra of the exact or truncated expansion being analyzed, but the alignment loss only constrains the portion visible through the selected probes, samples, and grid.

On hardware, estimating the full Pauli basis is expensive. Practical implementations may use a fixed subset of probes or randomized Pauli measurements, which gives a lower cost proxy for the same memory spectrum.

For any nonnegative spectrum \(R:\Omega_{\mathrm{grid}}\to\RR_{\ge0}\), the normalized spectrum used by the alignment loss is
\begin{equation}
\mathcal N_\delta R(\bar\omega_h)=
\frac{R(\bar\omega_h)+\delta}
{\sum_{g=1}^{N_\omega}R(\bar\omega_g)+N_\omega\delta}.
\end{equation}
The stop gradient in Eq.~\eqref{eq:align-loss} is applied only to the normalized prefix feature occupancy, not to the memory occupancy.

\newpage
\section{Limitations}
\label{app:lmt}

\paragraph{Spectral approximation and memory alignment.}
The spectral propagation result is exact when the composed memory state admits the finite coordinate Fourier expansion in Eq.~\eqref{eq:memory-fourier-main}. In implementations with sinusoidal positional encodings, nonlinear angle maps, finite sampling, or hardware estimates, this expansion is generally analyzed through a finite truncation or grid projection. The alignment objective in Section~\ref{sec:loss} inherits this approximation, since it matches the memory spectrum and prefix feature spectrum only on the chosen grid \(\Omega_{\mathrm{grid}}\) and only through the selected Pauli memory probes. Appendix~\ref{app:align-details} makes this proxy interpretation explicit. Frequencies outside the grid are not directly constrained. They can leak into nearby coefficients, alias under the sampled coordinate set, remain unresolved when they are too high frequency for the grid, or be suppressed by the ridge term in Eq.~\eqref{eq:fourier-ls-app}. Therefore \(\mathcal L_{\mathrm{align}}\) should be interpreted as a practical mechanism for shaping the resolved memory spectrum.

\paragraph{Hardware execution and measurement cost.}
SQMM also inherits practical limitations common to parameterized QML and recurrent quantum models. Pauli readout features are expectation values, so hardware execution requires repeated circuit runs with fixed parameters. For a band feature \(f_{i,a}(x)\), the circuit is executed up to band \(i\), \(O_{i,a}\) is measured on \(D_i\), and outcomes are averaged across shots. The measured state is not reused for later bands. Downstream features use separate executions that preserve the coherent memory path without conditioning on single shot outcomes, as formalized in Appendix~\ref{app:readout}. This creates a measurement overhead. Estimating the \(d_F\) readout observables at one coordinate uses \(\mathcal O(d_F/\mathrm{e}^2)\) shots for standard error \(\mathrm{e}\), so a pass over \(T\) coordinates uses \(\mathcal O(d_FT/\mathrm{e}^2)\) shots~\citep{knill2007optimal}. Full spectral alignment is more expensive and uses \(\mathcal O(4^mT(n-1)/\mathrm{e}^2)\) shots per training update for an \(m\) qubit memory, \(T\) sampled coordinates and \(n-1\) boundaries~\citep{kyrillidis2018provable}, unless subset probes or shadow based estimators are used~\citep{huang2020predicting}. In addition, noise affects both the memory evolution and the readout estimates. Appendix~\ref{app:noise} evaluates controlled circuit perturbations and observes smooth degradation, but this remains a simulator based study rather than a complete characterization of device noise, calibration error, decoherence, or measurement drift on real hardware.

\section{Further Experiments and Ablations}
\label{app:exp}

In this section, we further explore SQMM. All SQMM results use the model of Section~\ref{sec:exp} at the epochs and growth intervals of the corresponding main table, unless stated otherwise.

\FloatBarrier
\subsection{Bands \& Time}\label{app:exp:bands}
\begin{figure*}[h!]
\centering
\includegraphics[width=\textwidth]{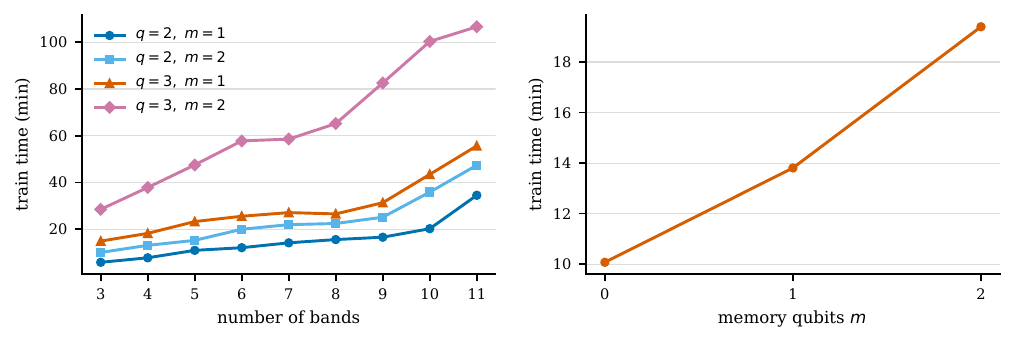}
\caption{\textbf{Mixed state scaling ablation on Cello.} Training time against the number of sequential bands (left) and the number of memory qubits (right). Each point is one run with 600 training epochs and a fixed depth per band.}
\label{fig:mixed-direct-scaling}
\end{figure*}
The setup for Figure~\ref{fig:mixed-direct-scaling} matches the cello data protocol in Section~\ref{exp:Classic_INR}.
The models were run without spectral loss, without classical preprocessing, and with linear postprocessing for 600 epochs.
Runs used five data reuploads per band.
The goal was to isolate how much the quantum component captures on its own and how circuit width and band count affect wall time and MSE.
Adding bands and memory qubits lowered the reconstruction error in these runs, at the training time cost shown in Figure~\ref{fig:mixed-direct-scaling}. Results reaffirm the known simulator cost of added qubits, where each additional qubit doubles the state dimension and increases time and memory requirements as demonstrated in the right graph.
Even a very small model with \(q=3\), \(m=1\), 11 bands, and 627 quantum parameters can take almost an hour to train on 1000 coordinates.
Bands in the 5 to 7 range gave the best tradeoff among runtime, parameter count, and reconstruction quality, which guided the later model search toward architectures with fewer than 7 bands.

\FloatBarrier
\subsection{Noise \& Hardware}\label{app:noise}

\begin{figure}[!htbp]
\centering
\includegraphics[width=\linewidth]{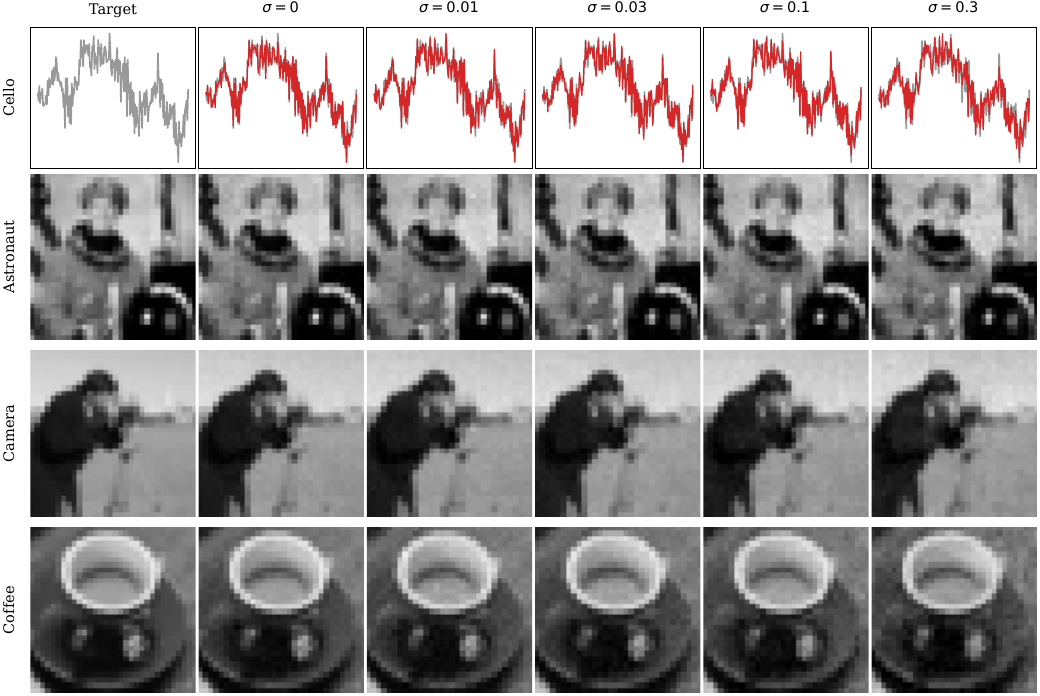}
\caption{Reconstructions under gate angle noise of SD \(\sigma\) in training and evaluation. Cello shows samples 400 to 600 with the target in grey and the prediction in red.}
\label{fig:images-noise-predictions}
\end{figure}

We perturb every gate angle with Gaussian noise of standard deviation \(\sigma\) during training and evaluation (Table~\ref{tab:noise-ablation}). The error grows smoothly with \(\sigma\) on the images and from \(\sigma=0.03\) on Cello, and the images are more sensitive to gate noise than Cello. Up to \(\sigma=0.03\) SQMM stays below every evaluated baseline on all four samples, and on Cello it does so at every \(\sigma\). Figure~\ref{fig:images-noise-predictions} shows the reconstructions at every \(\sigma\).

\begin{table}[!htbp]
\centering
\small
\setlength{\tabcolsep}{3pt}
\caption{Gaussian gate angle noise with SD \(\sigma\) in training and evaluation of the final model, MSE\(\downarrow\) \((\times 10^{-3})\), mean \(\pm\) SD. \(\sigma=0\) is the noiseless model.}
\label{tab:noise-ablation}
\begin{tabular}{lcccc}
\toprule
Perturbation & Cello & Astronaut & Camera & Coffee \\
\midrule
$\sigma=0$ & 0.48{\scriptsize$\pm$0.08} & 1.12{\scriptsize$\pm$0.10} & 0.38{\scriptsize$\pm$0.08} & 0.78{\scriptsize$\pm$0.05} \\
$\sigma=0.01$ & 0.46{\scriptsize$\pm$0.11} & 1.46{\scriptsize$\pm$0.11} & 0.44{\scriptsize$\pm$0.05} & 0.97{\scriptsize$\pm$0.32} \\
$\sigma=0.03$ & 0.56{\scriptsize$\pm$0.06} & 1.62{\scriptsize$\pm$0.14} & 0.68{\scriptsize$\pm$0.05} & 1.10{\scriptsize$\pm$0.16} \\
$\sigma=0.1$ & 0.73{\scriptsize$\pm$0.14} & 2.48{\scriptsize$\pm$0.56} & 1.03{\scriptsize$\pm$0.15} & 1.80{\scriptsize$\pm$0.36} \\
$\sigma=0.3$ & 1.13{\scriptsize$\pm$0.19} & 4.89{\scriptsize$\pm$0.58} & 2.38{\scriptsize$\pm$0.76} & 3.90{\scriptsize$\pm$1.00} \\
\bottomrule
\end{tabular}
\end{table}

\FloatBarrier
\subsection{Loss Extension}\label{app:loss}
\paragraph{1D rFFT.} For 1D uniformly sampled audio, Eq.~\eqref{eq:fourier-ls-main} with \(\varepsilon=0\) and \(\Omega_{\mathrm{grid}}\) the nonnegative discrete Fourier frequencies equals the real FFT scaled by \(1/T\), since the columns of \(\Phi\) are orthogonal. The feature and probe traces are real valued, so the negative frequency coefficients are the conjugates of the positive ones and carry no further information. The 1D real FFT is therefore an efficient implementation of the finite grid spectral projection on this grid.

\paragraph{Final output spectral control.}
The ablation in Table~\ref{tab:current-winner-memory-spectral-ex} also uses a control loss that applies the same spectral comparison directly to the final prediction. This loss is not the proposed memory alignment objective.
It does not inspect \(\sigma_M^{(i)}(x)\), does not use prefix occupancies, and does not provide an intermediate target for the memory lane.
It only asks the final predicted signal to have the same normalized spectral energy distribution as the target signal on a finite frequency grid.
For a generic nonnegative spectrum \(R\) on a finite grid \(\Omega\), write this normalization compactly as
\begin{equation}
\mathcal N_\delta R(\omega)
=
\frac{R(\omega)+\delta}
{\sum_{\omega'\in\Omega}R(\omega')+|\Omega|\delta}.
\label{eq:spectral-normalization-operator}
\end{equation}

Let \(x_1,\ldots,x_T\) be the sampled coordinates used in a training batch, and let
\[
\hat y_t=\hat y(x_t),
\qquad
y_t=y(x_t).
\]
For each output channel \(o\in\{1,\ldots,d_y\}\), estimate Fourier coefficients on a grid
\(\Omega_{\mathrm{out}}\) using the same FFT, DFT, or least squares estimator as above.
Write these coefficients as
\[
\widehat a_{\hat y,o}(\omega),
\qquad
\widehat a_{y,o}(\omega),
\qquad
\omega\in\Omega_{\mathrm{out}}.
\]
The final output spectral energies are
\begin{equation}
P_{\hat y}^{\mathrm{out}}(\omega)
=
\sum_{o=1}^{d_y}
\left|\widehat a_{\hat y,o}(\omega)\right|^2,
\qquad
P_{y}^{\mathrm{out}}(\omega)
=
\sum_{o=1}^{d_y}
\left|\widehat a_{y,o}(\omega)\right|^2.
\label{eq:output-spectrum-energy}
\end{equation}
After applying the same normalization operator as Eq.~\eqref{eq:align-normalization}, the output spectral control loss is
\begin{equation}
\mathcal L_{\mathrm{out}}
=
\sum_{\omega\in\Omega_{\mathrm{out}}}
\left(
\mathcal N_\delta P_{\hat y}^{\mathrm{out}}(\omega)
-
\operatorname{sg}\!\left[
\mathcal N_\delta P_y^{\mathrm{out}}(\omega)
\right]
\right)^2.
\label{eq:output-spectral-loss}
\end{equation}
The corresponding control objective is
\begin{equation}
\mathcal L
=
\mathcal L_{\mathrm{task}}
+
\lambda_{\mathrm{out}}\mathcal L_{\mathrm{out}}.
\label{eq:output-spectral-objective}
\end{equation}
The stop gradient is applied only to the target spectrum. Gradients flow through the final prediction \(\hat y\), so this control regularizes the predictor output itself. In contrast, Eq.~\eqref{eq:align-loss} shapes the propagated memory spectrum after each prefix and leaves the supervised task loss responsible for fitting the final signal values.
In the experiments, \(\mathcal L_{\mathrm{out}}\) uses \(\lambda_{\mathrm{out}}=0.1\).

\paragraph{No memory controls.} The no memory control of Table~\ref{tab:nomem} keeps the classical part of SQMM, including the projection that prepares the memory, and replaces the circuit by independent bands of four qubits. In every band the two data qubits are uploaded in every layer and read out through the same Pauli \(Z\) and \(ZZ\) features as in SQMM. The two further qubits are prepared once from the memory projection, take part in every entangling layer and are never measured. They start from \(\lvert 0\rangle\) in every band, so nothing passes between bands. The bands use \([1,2,3,3,4]\) layers, which gives 650 parameters against 642 for SQMM, and the control is trained with the protocol of Table~\ref{tab:classic-inr} without an auxiliary loss, since it has no shared memory to align. Table~\ref{tab:current-winner-memory-spectral-ex} repeats this control on all four samples as No memory, local qubits, with 662 parameters against 654 for SQMM on the images, and uses two further controls. No memory qubits removes the memory qubits and their couplings, leaving two qubits per band. No memory qubits with a wider decoder widens the decoder to 11 hidden units, which restores the parameter count but not the qubit count.

\paragraph{Mechanism ablation.} Table~\ref{tab:current-winner-memory-spectral-ex} extends the mechanism ablation of Table~\ref{tab:nomem} to all four samples. Removing the memory qubits gives the largest error on every sample, and a wider decoder at a similar parameter count recovers only part of the gap. The control with local qubits has a higher error than SQMM on Cello, Astronaut and Camera and the same error on Coffee. Read only memory lies between no memory and read and write memory. \(\mathcal L_{\mathrm{out}}\) leaves the error of the memory model nearly unchanged, while \(\mathcal L_{\mathrm{align}}\) gives the lowest error on every sample. The one memory qubit settings train SQMM with \(\mathcal L_{\mathrm{align}}\) and a single memory qubit. The first keeps the bands of SQMM, and the second adds one reupload to band 1 and two to each of bands 2 to 5, which matches the parameter count of SQMM. With one memory qubit the error rises on every sample. At the parameter count of SQMM the added reuploads close part of the gap, most on Cello where the two errors lie within one SD, and SQMM keeps the lowest error on every sample. The second memory qubit is therefore worth more than the same number of parameters spent on reuploads.

\begin{table*}[!htbp]
\centering
\small
\setlength{\tabcolsep}{3pt}
\caption{Mechanism and control ablation on the classic benchmark, MSE\(\downarrow\) \((\times 10^{-3})\), mean \(\pm\) SD. Params give the live parameters of each setting. No memory qubits removes the memory qubits and their couplings, and No memory, local qubits is the control of Table~\ref{tab:nomem}. Both one qubit memory rows use \(\mathcal L_{\mathrm{align}}\), and the added reuploads match the parameter count of SQMM. Best in \textbf{bold}.}
\label{tab:current-winner-memory-spectral-ex}
\begin{tabular}{lcccccc}
\toprule
Setting & Params 1D & Params 2D & Cello & Astronaut & Camera & Coffee \\
\midrule
No memory qubits & 534 & 544 & 0.82{\scriptsize$\pm$0.06} & 2.94{\scriptsize$\pm$0.54} & 0.89{\scriptsize$\pm$0.07} & 1.59{\scriptsize$\pm$0.42} \\
No memory qubits, wider decoder & 636 & 646 & 0.65{\scriptsize$\pm$0.08} & 1.94{\scriptsize$\pm$0.18} & 0.73{\scriptsize$\pm$0.20} & 1.33{\scriptsize$\pm$0.21} \\
No memory, local qubits & 650 & 662 & 0.53{\scriptsize$\pm$0.06} & 1.41{\scriptsize$\pm$0.18} & 0.47{\scriptsize$\pm$0.09} & 0.80{\scriptsize$\pm$0.06} \\
Read only memory & 626 & 638 & 0.61{\scriptsize$\pm$0.13} & 1.52{\scriptsize$\pm$0.09} & 0.48{\scriptsize$\pm$0.09} & 1.00{\scriptsize$\pm$0.21} \\
Read and write memory & 642 & 654 & 0.52{\scriptsize$\pm$0.08} & 1.23{\scriptsize$\pm$0.21} & 0.42{\scriptsize$\pm$0.06} & 0.88{\scriptsize$\pm$0.11} \\
Memory with $\mathcal L_{\mathrm{out}}$ & 642 & 654 & 0.54{\scriptsize$\pm$0.06} & 1.20{\scriptsize$\pm$0.15} & 0.42{\scriptsize$\pm$0.07} & 0.88{\scriptsize$\pm$0.08} \\
One qubit memory with $\mathcal L_{\mathrm{align}}$ & 588 & 599 & 0.66{\scriptsize$\pm$0.09} & 1.78{\scriptsize$\pm$0.20} & 0.59{\scriptsize$\pm$0.11} & 1.34{\scriptsize$\pm$0.11} \\
One qubit memory, more reuploads & 642 & 653 & 0.52{\scriptsize$\pm$0.05} & 1.38{\scriptsize$\pm$0.17} & 0.49{\scriptsize$\pm$0.08} & 1.21{\scriptsize$\pm$0.21} \\
Memory with $\mathcal L_{\mathrm{align}}$ (Ours) & 642 & 654 & \best{0.48{\scriptsize$\pm$0.08}} & \best{1.12{\scriptsize$\pm$0.10}} & \best{0.38{\scriptsize$\pm$0.08}} & \best{0.78{\scriptsize$\pm$0.05}} \\
\bottomrule
\end{tabular}
\end{table*}

\paragraph{Memory switched off at evaluation.} To test whether the trained model uses its memory, we take the five runs of Table~\ref{tab:classic-inr} and evaluate each again with every memory read and write coupling set to zero. The read and write gates are controlled rotations and \(XY\) gates, which reduce to the identity at zero angle, so the bands run exactly as trained while no information passes through the memory register. Nothing is retrained. Switching the memory off raises the error by a factor of \vMemOffCello{} on Cello, \vMemOffAstronaut{} on Astronaut, \vMemOffCamera{} on Camera and \vMemOffCoffee{} on Coffee (Table~\ref{tab:memory-off}), far above the error of a model trained without memory. The trained bands therefore rely on what the memory carries, and the memory register is not a component that the model learns to ignore.

\begin{table}[!htbp]
\centering
\small
\setlength{\tabcolsep}{3pt}
\caption{Memory switched off at evaluation on the classic benchmark, MSE\(\downarrow\) \((\times 10^{-3})\), mean \(\pm\) SD over the five runs of Table~\ref{tab:classic-inr}. The first two rows evaluate the same trained models, with every read and write coupling set to zero in the second. The last row is the model without memory qubits of Table~\ref{tab:current-winner-memory-spectral-ex}. \(\text{}^\star\) marks our model.}
\label{tab:memory-off}
\begin{tabular}{lcccc}
\toprule
Setting & Cello & Astronaut & Camera & Coffee \\
\midrule
SQMM$^\star$, memory on & 0.48{\scriptsize$\pm$0.08} & 1.12{\scriptsize$\pm$0.10} & 0.38{\scriptsize$\pm$0.08} & 0.78{\scriptsize$\pm$0.05} \\
Memory switched off at evaluation & 17.05{\scriptsize$\pm$4.65} & 442.40{\scriptsize$\pm$194.67} & 110.84{\scriptsize$\pm$37.98} & 113.69{\scriptsize$\pm$50.65} \\
\midrule
Trained without memory qubits & 0.82{\scriptsize$\pm$0.06} & 2.94{\scriptsize$\pm$0.54} & 0.89{\scriptsize$\pm$0.07} & 1.59{\scriptsize$\pm$0.42} \\
\bottomrule
\end{tabular}
\end{table}

\FloatBarrier
\paragraph{Memory anatomy.}\label{app:anatomy}
Memory is an input dependent reduced quantum state rather than a feature vector, so we resolve it at every band boundary along three axes in Table~\ref{tab:memory-anatomy}. Spectral alignment compares the memory occupancy spectrum with the accumulated prefix spectrum by cosine similarity and Jensen Shannon divergence. Coordinate spread is the variance of \(\langle P\rangle(x)\) across coordinates summed over the 15 non identity Paulis, which measures how far the state moves with the input. Write innovation is the norm of the Bloch vector that a write produces from a maximally mixed input, which isolates what the write injects from what memory retains. With \(\mathcal L_{\mathrm{align}}\), cosine agreement rises from the first to the last boundary while the coordinate spread decreases with depth. The write innovation is largest at the first boundary. Switching the memory off at evaluation raises the error of the trained Astronaut model by a factor of \vMemOffAstronaut{} (Table~\ref{tab:memory-off}).

\begin{table}[!htbp]
\centering
\small
\setlength{\tabcolsep}{3pt}
\caption{Memory anatomy on Astronaut at each band boundary with \(\mathcal L_{\mathrm{align}}\).}
\label{tab:memory-anatomy}
\begin{tabular}{lcccc}
\toprule
Boundary & Cosine$\uparrow$ & JS$\downarrow$ & Coordinate spread & Write innovation \\
\midrule
1 & 0.75 & 0.12 & 1.64 & 0.21 \\
2 & 0.82 & 0.11 & 1.26 & 0.15 \\
3 & 0.82 & 0.14 & 1.02 & 0.08 \\
4 & 0.82 & 0.17 & 0.80 & 0.09 \\
\bottomrule
\end{tabular}
\end{table}

\FloatBarrier
\subsection{Finite shot measurement}\label{app:finite}

\begin{figure}[!htbp]
\centering
\includegraphics[width=\linewidth]{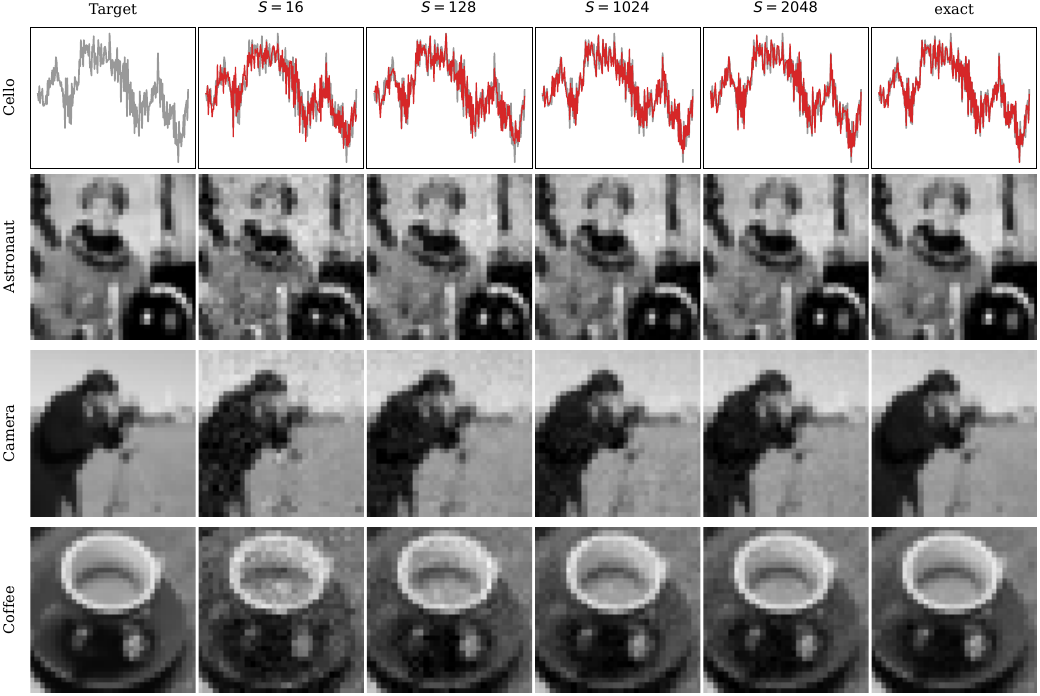}
\caption{Reconstructions with expectations estimated from \(S\) shots in training and evaluation. Cello shows samples 400 to 600 with the target in grey and the prediction in red.}
\label{fig:shot-reconstructions}
\end{figure}

Table~\ref{tab:cello-shot-learning-v2} trains and evaluates SQMM with every expectation estimated from \(S\) shots, on all four samples at every \(S\). On every sample the error decreases with \(S\), and on Cello at 2048 shots it is \vShotGapPct\,\% above exact evaluation. The images need more shots than Cello to approach exact evaluation. On Cello SQMM is below every evaluated baseline from 32 shots on, while on the images its error at 2048 shots is still above that of the strongest baselines. Figure~\ref{fig:shot-reconstructions} shows how the reconstructions change with the shot count.

\begin{table}[!htbp]
\centering
\small
\setlength{\tabcolsep}{3pt}
\caption{MSE\(\downarrow\) \((\times 10^{-3})\) with expectations estimated from \(S\) shots in training and evaluation, mean \(\pm\) SD.}
\label{tab:cello-shot-learning-v2}
\begin{tabular}{lcccc}
\toprule
Shots $S$ & Cello & Astronaut & Camera & Coffee \\
\midrule
16 & 3.65{\scriptsize$\pm$0.23} & 18.39{\scriptsize$\pm$1.76} & 7.64{\scriptsize$\pm$0.78} & 10.04{\scriptsize$\pm$0.45} \\
32 & 3.05{\scriptsize$\pm$0.12} & 12.24{\scriptsize$\pm$0.82} & 5.89{\scriptsize$\pm$0.84} & 8.15{\scriptsize$\pm$1.01} \\
64 & 2.30{\scriptsize$\pm$0.17} & 9.50{\scriptsize$\pm$0.70} & 4.07{\scriptsize$\pm$0.29} & 5.82{\scriptsize$\pm$0.43} \\
128 & 1.81{\scriptsize$\pm$0.14} & 7.86{\scriptsize$\pm$0.43} & 3.38{\scriptsize$\pm$0.26} & 4.61{\scriptsize$\pm$0.57} \\
256 & 1.39{\scriptsize$\pm$0.08} & 6.17{\scriptsize$\pm$0.57} & 2.63{\scriptsize$\pm$0.44} & 3.76{\scriptsize$\pm$1.14} \\
512 & 1.12{\scriptsize$\pm$0.07} & 4.70{\scriptsize$\pm$0.60} & 2.21{\scriptsize$\pm$0.38} & 3.00{\scriptsize$\pm$0.35} \\
1024 & 0.96{\scriptsize$\pm$0.09} & 3.73{\scriptsize$\pm$0.64} & 1.70{\scriptsize$\pm$0.19} & 2.22{\scriptsize$\pm$0.20} \\
2048 & 0.76{\scriptsize$\pm$0.06} & 2.86{\scriptsize$\pm$0.53} & 1.27{\scriptsize$\pm$0.16} & 1.79{\scriptsize$\pm$0.16} \\
exact & 0.48{\scriptsize$\pm$0.08} & 1.12{\scriptsize$\pm$0.10} & 0.38{\scriptsize$\pm$0.08} & 0.78{\scriptsize$\pm$0.05} \\
\bottomrule
\end{tabular}
\end{table}

\FloatBarrier
\subsection{Model configurations}\label{app:config}\label{app:baselines}

All experiments run on CPU nodes of a shared internal cluster, and the CPU type and memory of a node vary with the availability of the shared system. Table~\ref{tab:sqmm-config} lists the settings of SQMM and Table~\ref{tab:baseline-config} the configurations of the classical baselines. The same SQMM is used for every dataset, with the coordinate encoding and spectrum bins of the input dimension and three outputs for RGB images.

\begin{table}[!htbp]
\centering
\small
\caption{Settings of SQMM.}
\label{tab:sqmm-config}
\begin{tabular}{l>{\raggedright\arraybackslash}p{0.72\linewidth}}
\toprule
Component & Setting \\
\midrule
Bands & 5 bands of 2 data qubits with 2, 3, 4, 4 and 5 uploads \\
Encoding & the raw coordinate and 16 dyadic sinusoidal frequencies uploaded through \(R_Y\) in 1D, the raw coordinates and 8 consecutive frequencies per axis uploaded through \(R_Y\) and \(R_X\) with a shared bias in 2D \\
Memory & 2 qubits coupled to every data qubit, no entangler of its own, initialized by \(R_Y\) and \(R_X\) uploads of the encoding \\
Read & controlled \(R_Y\) before the first upload of every band, an extra controlled \(R_X\) in band 5 \\
Write & \(XY\) couplings in bands 1 to 4, none in band 5 \\
Readout & Pauli \(Z\) and \(ZZ\) expectations of every band, concatenated \\
Decoder & MLP with one hidden layer of width 5 and GELU, three outputs for RGB images \\
Training & Adam, learning rate 0.03, full batch, one update per epoch \\
Band growth & one band per growth interval of 25 updates for 300 epochs and 50 updates for 600 epochs, with a fade in of 5 updates \\
\(\mathcal L_{\mathrm{align}}\) & \(\lambda_{\mathrm{align}}=3\), \(\delta=10^{-8}\), boundaries 1 to 3, switched off after four growth intervals \\
Bins & all \(\lfloor T/2\rfloor+1\) nonnegative real FFT bins in 1D (501 on Cello, 513 on the audio clips), the \(32\times32\) lowest signed frequencies of the grid FFT in 2D (\(N_\omega=1{,}024\) at every image size) \\
\bottomrule
\end{tabular}
\end{table}

\textbf{Spectral memory alignment.} The feature spectrum uses the Pauli \(Z\) and \(ZZ\) features of every active band, and the memory spectrum uses the \(4^m\) full Pauli signals \(\chi^M_{i,s}(x)\) of the memory state after band \(i\), which are available in simulation. All coordinates lie on regular grids, so Eq.~\eqref{eq:fourier-ls-app} is used with \(\varepsilon=0\) and DFT frequencies, where it equals the discrete Fourier transform scaled by \(1/T\). The implementation uses the orthonormal FFT, which multiplies both spectra by \(T\). Energies are summed over the feature components or Pauli words of a band. The prefix target \(P_i^F\) is the cumulative sum of the band energies of bands \(1\) to \(i\), normalized after summation, and \(P_i^M\) is normalized in the same way, so the factors \(2^{-m}\) and \(T\) cancel up to a rescaling of \(\delta\) to \(\delta/T\) for the feature spectrum and to \(\delta/(2^mT)\) for the memory spectrum. \(\mathcal L_{\mathrm{align}}\) is the squared difference of the normalized spectra averaged over the active boundaries and bins, without frequency weighting. Boundary \(i\) enters the loss when band \(i+1\) is added, and the loss is switched off when the fifth band is added, so boundary 4, which is read only by the last band, is not supervised. The target is detached, and gradients of the memory side reach every parameter the memory state depends on.

\textbf{Classical baselines.} Every classical baseline has one configuration for audio (1D), fitted on Cello, GTZAN, UrbanSound8K and VCTK, and one for images (2D), fitted on Astronaut, Camera, Coffee, CIFAR10, Kodak and DIV2K with three outputs for RGB. A configuration fixes the hidden width, the number of hidden layers and the learning rate. The 1D shape is the shape selected for the audio comparison and the 2D shape the shape selected for the CIFAR10 comparison, both from every width and depth shape whose parameter count falls inside the budget. The learning rate is selected from \([3\cdot10^{-2},10^{-3}]\), by the lowest geometric mean of the best epoch MSE over Cello and the audio corpora in 1D and over Astronaut, Camera, Coffee and CIFAR10 in 2D. Kodak and DIV2K reuse the selected 2D configuration. All baselines are trained with Adam, full batch updates and no learning rate schedule on the MSE of targets in \([-1,1]\), for the epochs of the corresponding SQMM experiment, and are evaluated with their best epoch weights.

Every other setting keeps the default of the official implementation. Two baselines change a default. MSA-INR uses a signed input domain and, in 2D, a frequency comb rescaled to the Nyquist band of the \(32\times32\) grid. The published CAFE encoder of 88 random Fourier features and Chebyshev order 30 admits a single hidden unit within the budget, so we also report CAFE (enc.), which scales the encoder down to 5 features and order 2 in 1D and 21 features and order 7 in 2D.

\begin{table}[!htbp]
\centering
\small
\setlength{\tabcolsep}{3pt}
\caption{Configurations of the classical baselines. Shape is hidden width \(\times\) hidden layers. TUNER has no 1D configuration since its initialization requires 2D coordinates.}
\label{tab:baseline-config}
\begin{tabular}{lcccc}
\toprule
 & \multicolumn{2}{c}{1D (audio)} & \multicolumn{2}{c}{2D (images)} \\
\cmidrule(lr){2-3}\cmidrule(lr){4-5}
Model & Shape & lr & Shape & lr \\
\midrule
FINER & \(10\times6\) & \(10^{-3}\) & \(24\times1\) & \(3\cdot10^{-3}\) \\
SIREN & \(14\times3\) & \(10^{-3}\) & \(14\times3\) & \(3\cdot10^{-3}\) \\
NeRF & \(17\times1\) & \(10^{-2}\) & \(12\times3\) & \(3\cdot10^{-2}\) \\
KAN & \(2\times4\) & \(3\cdot10^{-2}\) & \(2\times4\) & \(10^{-2}\) \\
WIRE & \(25\times2\) & \(3\cdot10^{-3}\) & \(25\times2\) & \(3\cdot10^{-3}\) \\
R+RFF & \(10\times5\) & \(3\cdot10^{-3}\) & \(24\times1\) & \(3\cdot10^{-2}\) \\
ReLU & \(11\times4\) & \(3\cdot10^{-3}\) & \(11\times4\) & \(10^{-2}\) \\
Tanh & \(16\times2\) & \(3\cdot10^{-2}\) & \(11\times4\) & \(3\cdot10^{-2}\) \\
IGA & \(14\times3\) & \(10^{-3}\) & \(14\times3\) & \(3\cdot10^{-3}\) \\
FLAIR & \(10\times4\) & \(3\cdot10^{-3}\) & \(17\times2\) & \(3\cdot10^{-2}\) \\
TUNER &  &  & \(25\times1\) & \(3\cdot10^{-3}\) \\
VI3NR & \(10\times6\) & \(3\cdot10^{-2}\) & \(10\times6\) & \(3\cdot10^{-2}\) \\
MSA-INR & \(24\times1\) & \(3\cdot10^{-2}\) & \(24\times1\) & \(3\cdot10^{-2}\) \\
CAFE (enc.) & \(13\times1\) & \(3\cdot10^{-2}\) & \(4\times1\) & \(2\cdot10^{-2}\) \\
CAFE & \(1\times1\) & \(2\cdot10^{-2}\) & \(1\times1\) & \(10^{-2}\) \\
\bottomrule
\end{tabular}
\end{table}

\textbf{Quantum baselines.} QIREN uses its published hybrid architecture of width 8 with 2 hybrid layers and CZ entanglers, with spectrum layer 2 (649 parameters in 1D, 657 in 2D), and QIREN+ uses spectrum layer 3 (793 in 1D, 801 in 2D), and learning rate \(0.03\). QVF uses its published 671 parameter image model with 6 embedding frequencies, a hidden width of 10, one repetition and learning rate \(0.01\).

\FloatBarrier
\subsection{Base benchmark best runs}\label{app:classic-bestrun}

Table~\ref{tab:classic-bestrun} reports the base benchmark of Table~\ref{tab:classic-inr} in the best run convention of QIREN, with all evaluated baselines and QIREN+, which adds one spectrum layer to QIREN at a larger parameter count. SQMM has the lowest best run error on every sample.

\begin{table*}[!htbp]
\centering
\small
\setlength{\tabcolsep}{3pt}
\caption{Classic INR benchmark in the best run convention of QIREN, MSE\(\downarrow\) \((\times 10^{-3})\) of the best single run. The SQMM rows take the best of the five runs in Table~\ref{tab:classic-inr}. QIREN+ is QIREN with one extra spectrum layer. \(\text{}^\star\) marks our model. Best in \textbf{bold}.}
\label{tab:classic-bestrun}
\begin{tabular}{lcccccc}
\toprule
Model & Params 1D & Params 2D & Cello & Astronaut & Camera & Coffee \\
\midrule
SQMM$^\star$ & 642 & 654 & \best{0.38} & \best{1.01} & \best{0.28} & \best{0.73} \\
SQMM$^\star_{2048}$ & 642 & 654 & 0.71 & 2.36 & 1.09 & 1.57 \\
\midrule
QIREN & 649 & 657 & 6.25 & 5.34 & 1.80 & 1.73 \\
QVF &  & 671 &  & 11.00 & 3.22 & 7.76 \\
QIREN+ & 793 & 801 & 5.66 & 3.17 & 1.39 & 1.31 \\
FINER & 691 & 697 & 3.32 & 2.49 & 1.25 & 1.54 \\
SIREN & 673 & 687 & 4.93 & 1.82 & 0.44 & 1.96 \\
NeRF & 698 & 997 & 4.42 & 4.52 & 1.16 & 2.82 \\
KAN & 702 & 1222 & 3.07 & 43.40 & 21.02 & 21.10 \\
WIRE & 1294 & 1311 & 4.81 & 19.19 & 26.41 & 23.89 \\
R+RFF & 661 & 673 & 5.45 & 4.23 & 2.50 & 3.61 \\
ReLU & 672 & 683 & 7.65 & 11.36 & 3.15 & 4.21 \\
Tanh & 689 & 683 & 14.99 & 17.79 & 7.32 & 12.56 \\
IGA & 673 & 687 & 4.84 & 1.38 & 0.59 & 1.11 \\
FLAIR & 703 & 687 & 4.69 & 4.40 & 2.69 & 2.64 \\
TUNER &  & 701 &  & 4.73 & 5.38 & 3.78 \\
VI3NR & 691 & 701 & 7.78 & 6.24 & 3.22 & 3.76 \\
MSA-INR & 673 & 697 & 6.37 & 1.58 & 0.99 & 1.08 \\
CAFE (enc.) & 703 & 709 & 2.81 & 3.42 & 1.01 & 3.50 \\
CAFE & 625 & 715 & 11.52 & 30.97 & 36.94 & 19.73 \\
\bottomrule
\end{tabular}
\end{table*}

\FloatBarrier
\subsection{Audio results}\label{app:audio-metrics}

Table~\ref{tab:audio-full} compares all models on the three audio corpora. SQMM has the lowest error on every corpus. FINER and CAFE (enc.) are the closest baselines, while QIREN, QIREN+ and MSA-INR stay close to an order of magnitude above SQMM. Table~\ref{tab:audio-metrics} reports further SQMM metrics. Speech (VCTK) is reconstructed with the lowest MSE and the highest PSNR, and music (GTZAN) has the highest log spectral distance.

\begin{table}[!htbp]
\centering
\small
\setlength{\tabcolsep}{3pt}
\caption{Audio reconstruction MSE\(\downarrow\) \((\times 10^{-3})\) for all models, mean \(\pm\) SD. QIREN+ adds one spectrum layer to QIREN. \(\text{}^\star\) marks our model. Best in \textbf{bold}.}
\label{tab:audio-full}
\begin{tabular}{lcccc}
\toprule
Model & Params & GTZAN & US8K & VCTK \\
\midrule
SQMM$^\star$ & 642 & \best{4.71{\scriptsize$\pm$0.17}} & \best{6.11{\scriptsize$\pm$0.27}} & \best{1.00{\scriptsize$\pm$0.03}} \\
\midrule
QIREN+ & 793 & 40.92{\scriptsize$\pm$1.15} & 50.18{\scriptsize$\pm$0.34} & 9.73{\scriptsize$\pm$0.76} \\
QIREN & 649 & 44.19{\scriptsize$\pm$1.88} & 51.83{\scriptsize$\pm$0.86} & 10.70{\scriptsize$\pm$0.10} \\
FINER & 691 & 5.14{\scriptsize$\pm$1.06} & 8.56{\scriptsize$\pm$2.35} & 1.66{\scriptsize$\pm$0.21} \\
SIREN & 673 & 8.65{\scriptsize$\pm$0.61} & 15.55{\scriptsize$\pm$1.49} & 2.98{\scriptsize$\pm$0.15} \\
NeRF & 698 & 8.89{\scriptsize$\pm$0.44} & 13.96{\scriptsize$\pm$0.72} & 2.59{\scriptsize$\pm$0.15} \\
KAN & 702 & 17.26{\scriptsize$\pm$0.37} & 18.48{\scriptsize$\pm$0.38} & 3.75{\scriptsize$\pm$0.12} \\
WIRE & 1294 & 25.64{\scriptsize$\pm$12.78} & 34.37{\scriptsize$\pm$7.31} & 5.51{\scriptsize$\pm$2.73} \\
R+RFF & 661 & 37.53{\scriptsize$\pm$7.30} & 48.50{\scriptsize$\pm$4.40} & 7.47{\scriptsize$\pm$1.54} \\
ReLU & 672 & 58.90{\scriptsize$\pm$4.77} & 58.16{\scriptsize$\pm$1.37} & 11.41{\scriptsize$\pm$0.77} \\
Tanh & 689 & 92.35{\scriptsize$\pm$0.35} & 70.14{\scriptsize$\pm$0.73} & 19.29{\scriptsize$\pm$0.39} \\
IGA & 673 & 8.45{\scriptsize$\pm$0.56} & 15.47{\scriptsize$\pm$1.22} & 2.69{\scriptsize$\pm$0.13} \\
FLAIR & 703 & 25.03{\scriptsize$\pm$3.93} & 39.47{\scriptsize$\pm$3.75} & 4.87{\scriptsize$\pm$0.54} \\
VI3NR & 691 & 50.98{\scriptsize$\pm$3.19} & 56.52{\scriptsize$\pm$1.69} & 11.14{\scriptsize$\pm$0.42} \\
MSA-INR & 673 & 42.11{\scriptsize$\pm$3.39} & 52.21{\scriptsize$\pm$1.55} & 8.51{\scriptsize$\pm$0.90} \\
CAFE (enc.) & 703 & 6.42{\scriptsize$\pm$1.34} & 10.83{\scriptsize$\pm$3.75} & 2.25{\scriptsize$\pm$0.34} \\
CAFE & 625 & 32.75{\scriptsize$\pm$34.02} & 36.96{\scriptsize$\pm$28.28} & 28.75{\scriptsize$\pm$39.99} \\
\bottomrule
\end{tabular}
\end{table}

\begin{table}[!htbp]
\centering
\small
\setlength{\tabcolsep}{3pt}
\caption{SQMM audio metrics per corpus, mean \(\pm\) SD.}
\label{tab:audio-metrics}
\begin{tabular}{lccc}
\toprule
Corpus & MSE$\downarrow$ (\(\times10^{-3}\)) & PSNR$\uparrow$ (dB) & LSD$\downarrow$ \\
\midrule
GTZAN & 4.71{\scriptsize$\pm$0.17} & 24.55{\scriptsize$\pm$0.17} & 7.10{\scriptsize$\pm$0.10} \\
US8K & 6.11{\scriptsize$\pm$0.27} & 22.92{\scriptsize$\pm$0.22} & 5.38{\scriptsize$\pm$0.08} \\
VCTK & 1.00{\scriptsize$\pm$0.03} & 33.94{\scriptsize$\pm$0.17} & 5.57{\scriptsize$\pm$0.06} \\
\bottomrule
\end{tabular}
\end{table}

\FloatBarrier
\subsection{Kodak and DIV2K}\label{app:ladder}

\begin{figure}[!htbp]
\centering
\includegraphics[width=0.55\linewidth]{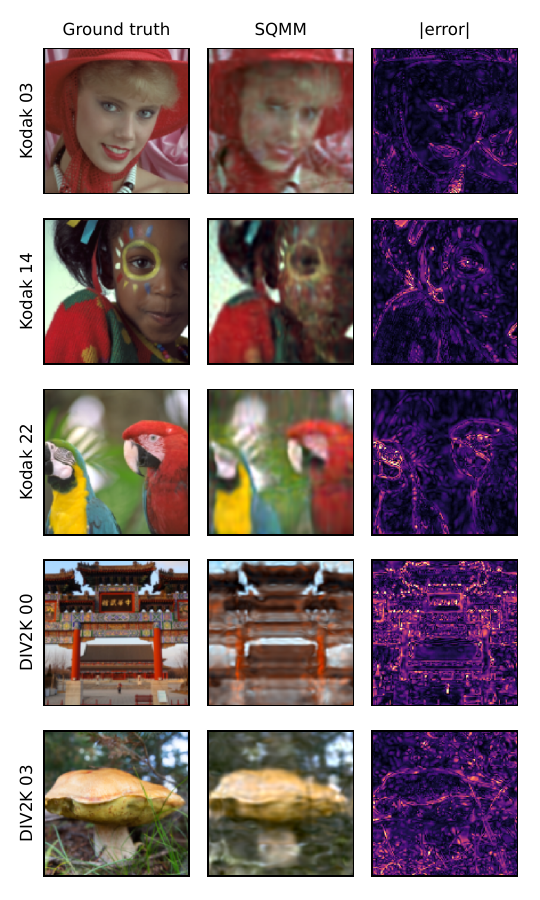}
\caption{SQMM reconstructions of Kodak and DIV2K images at \(128\times128\). Columns show the ground truth, the SQMM reconstruction, and the absolute error.}
\label{fig:ladder-grid-128}
\end{figure}

Table~\ref{tab:kodak-native-full} compares all models on the 24 Kodak images at native resolution, and Tables~\ref{tab:ladder-kodak} and~\ref{tab:ladder-div2k} give the resolution ladders. SQMM has the lowest error at every resolution on both datasets, with the fewest parameters. The error of SQMM and of most baselines grows with resolution, since the parameter budget is fixed while the number of pixels grows. MSA-INR is the closest baseline up to \(64^2\), and SIREN or IGA at larger sizes, and the relative gap between SQMM and the closest baseline narrows as the resolution grows. All 2D configurations are selected on \(32\times32\) images, and WIRE and TUNER degrade sharply from \(64^2\) on. In Figure~\ref{fig:ladder-grid-128} the remaining SQMM error concentrates on edges and fine textures, while smooth regions are reproduced closely.

\begin{table}[!htbp]
\centering
\small
\setlength{\tabcolsep}{3pt}
\caption{Kodak at native resolution (\(768\times512\), 24 images) for all models, MSE \((\times 10^{-3})\), mean \(\pm\) SD. \(\text{}^\star\) marks our model. Best in \textbf{bold}.}
\label{tab:kodak-native-full}
\begin{tabular}{lcc}
\toprule
Model & Params & MSE$\downarrow$ (\(\times10^{-3}\)) \\
\midrule
SQMM$^\star$ & 666 & \best{32.08{\scriptsize$\pm$0.15}} \\
\midrule
FINER & 747 & 39.13{\scriptsize$\pm$0.59} \\
SIREN & 717 & 34.44{\scriptsize$\pm$0.61} \\
NeRF & 1023 & 34.79{\scriptsize$\pm$0.40} \\
KAN & 1274 & 54.10{\scriptsize$\pm$1.35} \\
WIRE & 1383 & 47.58{\scriptsize$\pm$3.21} \\
R+RFF & 723 & 38.54{\scriptsize$\pm$0.64} \\
ReLU & 707 & 41.51{\scriptsize$\pm$1.37} \\
Tanh & 707 & 45.00{\scriptsize$\pm$0.35} \\
IGA & 717 & 34.04{\scriptsize$\pm$0.56} \\
FLAIR & 723 & 39.04{\scriptsize$\pm$0.39} \\
TUNER & 753 & 118.47{\scriptsize$\pm$0.99} \\
VI3NR & 723 & 41.73{\scriptsize$\pm$1.00} \\
MSA-INR & 747 & 35.64{\scriptsize$\pm$0.26} \\
CAFE (enc.) & 719 & 48.45{\scriptsize$\pm$2.09} \\
CAFE & 719 & 63.64{\scriptsize$\pm$0.51} \\
\bottomrule
\end{tabular}
\end{table}

\begin{table*}[!htbp]
\centering
\small
\setlength{\tabcolsep}{3pt}
\caption{Kodak resolution ladder (24 images, RGB), MSE\(\downarrow\) \((\times 10^{-3})\), mean \(\pm\) SD. \(\text{}^\star\) marks our model. Best in \textbf{bold}.}
\label{tab:ladder-kodak}
\begin{tabular}{lcccccc}
\toprule
Model & Params & $32^2$ & $64^2$ & $128^2$ & $256^2$ & native $768{\times}512$ \\
\midrule
SQMM$^\star$ & 666 & \best{1.58{\scriptsize$\pm$0.10}} & \best{5.43{\scriptsize$\pm$0.03}} & \best{11.54{\scriptsize$\pm$0.05}} & \best{18.52{\scriptsize$\pm$0.04}} & \best{32.08{\scriptsize$\pm$0.15}} \\
\midrule
FINER & 747 & 5.82{\scriptsize$\pm$0.36} & 10.18{\scriptsize$\pm$0.26} & 17.32{\scriptsize$\pm$0.24} & 25.03{\scriptsize$\pm$0.34} & 39.13{\scriptsize$\pm$0.59} \\
SIREN & 717 & 2.70{\scriptsize$\pm$0.33} & 6.89{\scriptsize$\pm$0.26} & 13.18{\scriptsize$\pm$0.47} & 20.57{\scriptsize$\pm$0.55} & 34.44{\scriptsize$\pm$0.61} \\
NeRF & 1023 & 3.80{\scriptsize$\pm$0.42} & 8.76{\scriptsize$\pm$0.68} & 14.94{\scriptsize$\pm$0.26} & 21.57{\scriptsize$\pm$0.23} & 34.79{\scriptsize$\pm$0.40} \\
KAN & 1274 & 19.49{\scriptsize$\pm$1.72} & 25.53{\scriptsize$\pm$1.26} & 34.69{\scriptsize$\pm$1.71} & 42.10{\scriptsize$\pm$2.41} & 54.10{\scriptsize$\pm$1.35} \\
WIRE & 1383 & 19.27{\scriptsize$\pm$1.68} & 52.10{\scriptsize$\pm$3.73} & 53.46{\scriptsize$\pm$9.73} & 37.37{\scriptsize$\pm$3.40} & 47.58{\scriptsize$\pm$3.21} \\
R+RFF & 723 & 3.35{\scriptsize$\pm$0.07} & 9.47{\scriptsize$\pm$0.18} & 16.89{\scriptsize$\pm$0.29} & 24.53{\scriptsize$\pm$0.34} & 38.54{\scriptsize$\pm$0.64} \\
ReLU & 707 & 5.46{\scriptsize$\pm$0.75} & 12.11{\scriptsize$\pm$0.95} & 19.81{\scriptsize$\pm$0.93} & 27.44{\scriptsize$\pm$1.07} & 41.51{\scriptsize$\pm$1.37} \\
Tanh & 707 & 7.17{\scriptsize$\pm$0.29} & 14.42{\scriptsize$\pm$0.39} & 22.68{\scriptsize$\pm$0.33} & 30.46{\scriptsize$\pm$0.45} & 45.00{\scriptsize$\pm$0.35} \\
IGA & 717 & 3.11{\scriptsize$\pm$0.34} & 7.06{\scriptsize$\pm$0.25} & 13.65{\scriptsize$\pm$0.36} & 20.94{\scriptsize$\pm$0.54} & 34.04{\scriptsize$\pm$0.56} \\
FLAIR & 723 & 3.48{\scriptsize$\pm$0.08} & 9.57{\scriptsize$\pm$0.46} & 17.57{\scriptsize$\pm$0.28} & 25.37{\scriptsize$\pm$0.14} & 39.04{\scriptsize$\pm$0.39} \\
TUNER & 753 & 11.31{\scriptsize$\pm$0.24} & 62.25{\scriptsize$\pm$0.58} & 93.88{\scriptsize$\pm$0.41} & 104.37{\scriptsize$\pm$0.26} & 118.47{\scriptsize$\pm$0.99} \\
VI3NR & 723 & 5.91{\scriptsize$\pm$1.01} & 12.63{\scriptsize$\pm$1.51} & 20.59{\scriptsize$\pm$1.94} & 28.11{\scriptsize$\pm$1.87} & 41.73{\scriptsize$\pm$1.00} \\
MSA-INR & 747 & 1.85{\scriptsize$\pm$0.06} & 6.84{\scriptsize$\pm$0.06} & 14.02{\scriptsize$\pm$0.05} & 21.54{\scriptsize$\pm$0.04} & 35.64{\scriptsize$\pm$0.26} \\
CAFE (enc.) & 719 & 7.50{\scriptsize$\pm$1.90} & 16.21{\scriptsize$\pm$3.13} & 26.82{\scriptsize$\pm$1.80} & 35.03{\scriptsize$\pm$2.60} & 48.45{\scriptsize$\pm$2.09} \\
CAFE & 719 & 26.90{\scriptsize$\pm$7.16} & 41.96{\scriptsize$\pm$12.87} & 44.68{\scriptsize$\pm$1.65} & 52.47{\scriptsize$\pm$0.42} & 63.64{\scriptsize$\pm$0.51} \\
\bottomrule
\end{tabular}
\end{table*}

\begin{table*}[!htbp]
\centering
\small
\setlength{\tabcolsep}{3pt}
\caption{DIV2K resolution ladder (10 images, RGB), MSE\(\downarrow\) \((\times 10^{-3})\), mean \(\pm\) SD. The largest size is the \(512^2\) prepared image. \(\text{}^\star\) marks our model. Best in \textbf{bold}.}
\label{tab:ladder-div2k}
\begin{tabular}{lcccccc}
\toprule
Model & Params & $32^2$ & $64^2$ & $128^2$ & $256^2$ & full $512^2$ \\
\midrule
SQMM$^\star$ & 666 & \best{3.13{\scriptsize$\pm$0.08}} & \best{11.13{\scriptsize$\pm$0.06}} & \best{23.16{\scriptsize$\pm$0.34}} & \best{35.15{\scriptsize$\pm$0.43}} & \best{54.13{\scriptsize$\pm$0.44}} \\
\midrule
FINER & 747 & 8.59{\scriptsize$\pm$0.56} & 19.16{\scriptsize$\pm$0.35} & 32.75{\scriptsize$\pm$0.69} & 45.71{\scriptsize$\pm$0.65} & 65.18{\scriptsize$\pm$0.65} \\
SIREN & 717 & 5.22{\scriptsize$\pm$0.74} & 14.37{\scriptsize$\pm$0.75} & 27.07{\scriptsize$\pm$1.02} & 39.34{\scriptsize$\pm$0.74} & 58.41{\scriptsize$\pm$1.14} \\
NeRF & 1023 & 6.91{\scriptsize$\pm$0.20} & 18.57{\scriptsize$\pm$0.66} & 30.87{\scriptsize$\pm$0.38} & 42.68{\scriptsize$\pm$1.26} & 61.08{\scriptsize$\pm$0.53} \\
KAN & 1274 & 28.79{\scriptsize$\pm$1.74} & 43.67{\scriptsize$\pm$2.16} & 60.42{\scriptsize$\pm$2.71} & 74.62{\scriptsize$\pm$0.54} & 93.99{\scriptsize$\pm$2.11} \\
WIRE & 1383 & 27.27{\scriptsize$\pm$2.02} & 81.35{\scriptsize$\pm$4.27} & 90.97{\scriptsize$\pm$8.55} & 66.31{\scriptsize$\pm$3.86} & 80.39{\scriptsize$\pm$2.91} \\
R+RFF & 723 & 6.12{\scriptsize$\pm$0.22} & 18.42{\scriptsize$\pm$0.41} & 32.63{\scriptsize$\pm$0.76} & 45.80{\scriptsize$\pm$0.94} & 65.14{\scriptsize$\pm$0.85} \\
ReLU & 707 & 12.08{\scriptsize$\pm$1.70} & 27.58{\scriptsize$\pm$3.08} & 42.66{\scriptsize$\pm$2.66} & 56.28{\scriptsize$\pm$2.63} & 75.50{\scriptsize$\pm$3.10} \\
Tanh & 707 & 16.12{\scriptsize$\pm$0.70} & 32.78{\scriptsize$\pm$1.19} & 49.41{\scriptsize$\pm$0.73} & 62.30{\scriptsize$\pm$0.85} & 82.72{\scriptsize$\pm$1.97} \\
IGA & 717 & 5.34{\scriptsize$\pm$0.55} & 15.22{\scriptsize$\pm$0.60} & 28.11{\scriptsize$\pm$0.98} & 40.02{\scriptsize$\pm$0.40} & 58.40{\scriptsize$\pm$0.39} \\
FLAIR & 723 & 7.06{\scriptsize$\pm$0.13} & 19.83{\scriptsize$\pm$0.28} & 35.18{\scriptsize$\pm$0.30} & 48.18{\scriptsize$\pm$0.70} & 68.11{\scriptsize$\pm$1.31} \\
TUNER & 753 & 16.02{\scriptsize$\pm$0.25} & 95.56{\scriptsize$\pm$0.92} & 147.69{\scriptsize$\pm$0.30} & 164.13{\scriptsize$\pm$1.04} & 182.94{\scriptsize$\pm$1.22} \\
VI3NR & 723 & 10.95{\scriptsize$\pm$1.39} & 24.18{\scriptsize$\pm$2.15} & 39.20{\scriptsize$\pm$1.88} & 52.54{\scriptsize$\pm$2.25} & 72.22{\scriptsize$\pm$2.38} \\
MSA-INR & 747 & 3.58{\scriptsize$\pm$0.09} & 13.66{\scriptsize$\pm$0.11} & 27.35{\scriptsize$\pm$0.41} & 39.84{\scriptsize$\pm$0.31} & 59.10{\scriptsize$\pm$0.46} \\
CAFE (enc.) & 719 & 11.43{\scriptsize$\pm$0.73} & 33.81{\scriptsize$\pm$4.74} & 53.77{\scriptsize$\pm$1.85} & 67.65{\scriptsize$\pm$3.37} & 87.98{\scriptsize$\pm$3.18} \\
CAFE & 719 & 41.56{\scriptsize$\pm$8.76} & 70.45{\scriptsize$\pm$23.49} & 74.60{\scriptsize$\pm$1.70} & 88.81{\scriptsize$\pm$2.42} & 107.59{\scriptsize$\pm$2.59} \\
\bottomrule
\end{tabular}
\end{table*}

\FloatBarrier
\subsection{Ansatz Diagram}

\begin{figure}[!htbp]
\centering
\includegraphics[width=\linewidth]{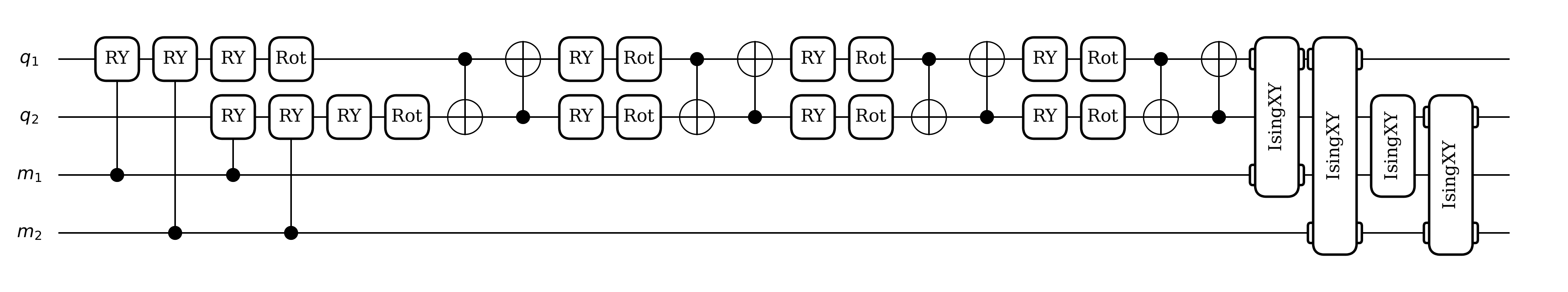}
\caption{Band 3 of SQMM, recorded from the trained Cello model. Controlled \(R_Y\) reads from the memory qubits \(m_1, m_2\) precede the four \(R_Y\) uploads on the data qubits \(q_1, q_2\), each followed by an entangling layer, and \(XY\) couplings write to the memory at the end of the band.}
\label{fig:cello_ansatz_pennylane}
\end{figure}

Figure~\ref{fig:cello_ansatz_pennylane} draws one band of the trained Cello model gate by gate, in the read, upload and write order of Section~\ref{sec:method}.

\FloatBarrier
\subsection{CIFAR10 image reconstruction}

\begin{figure*}[!htbp]
\centering
\includegraphics[width=\textwidth]{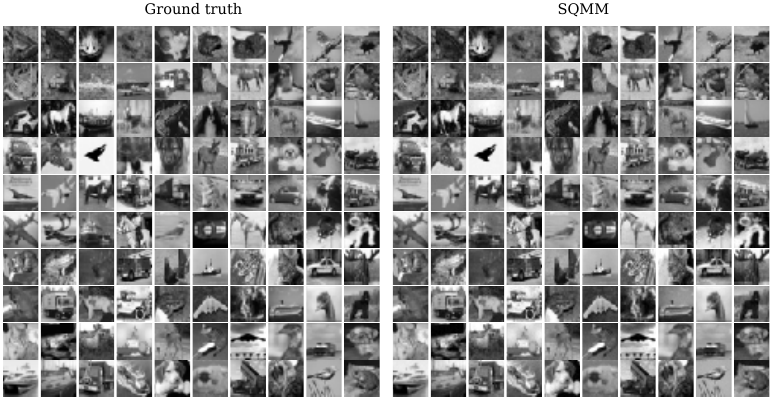}
\caption{The 100 grayscale CIFAR10 images (left) and their SQMM reconstructions (right).}
\label{fig:cifar10-gray-10x10-v2}
\end{figure*}

Table~\ref{tab:cifar-full} compares all models on the 100 grayscale CIFAR10 images. SQMM is best on all three metrics. MSA-INR, IGA and SIREN are the closest baselines, with about twice the MSE and more than twice the LPIPS of SQMM. Figure~\ref{fig:cifar10-gray-10x10-v2} shows the images with their SQMM reconstructions, which stay visually close to the ground truth at this resolution.

\begin{table}[!htbp]
\centering
\small
\setlength{\tabcolsep}{3pt}
\caption{Grayscale CIFAR10 metrics for all models, mean \(\pm\) SD. MSE is on the \([-1,1]\) range, SSIM and LPIPS on unit range images. QIREN+ adds one spectrum layer to QIREN. \(\text{}^\star\) marks our model. Best in \textbf{bold}.}
\label{tab:cifar-full}
\begin{tabular}{lcccc}
\toprule
Model & Params & SSIM$\uparrow$ (\(\times10^{-2}\)) & LPIPS$\downarrow$ (\(\times10^{-4}\)) & MSE$\downarrow$ (\(\times10^{-3}\)) \\
\midrule
SQMM$^\star$ & 654 & \best{99.21{\scriptsize$\pm$0.03}} & \best{3.44{\scriptsize$\pm$0.25}} & \best{0.46{\scriptsize$\pm$0.02}} \\
\midrule
QIREN+ & 801 & 97.69{\scriptsize$\pm$0.19} & 30.41{\scriptsize$\pm$2.98} & 1.67{\scriptsize$\pm$0.13} \\
QVF & 671 & 92.90{\scriptsize$\pm$0.23} & 104.84{\scriptsize$\pm$8.38} & 5.12{\scriptsize$\pm$0.17} \\
FINER & 697 & 97.43{\scriptsize$\pm$0.39} & 14.17{\scriptsize$\pm$3.69} & 1.45{\scriptsize$\pm$0.21} \\
SIREN & 687 & 98.37{\scriptsize$\pm$0.32} & 8.22{\scriptsize$\pm$1.89} & 0.98{\scriptsize$\pm$0.21} \\
WIRE & 1311 & 83.21{\scriptsize$\pm$2.51} & 306.79{\scriptsize$\pm$62.91} & 18.23{\scriptsize$\pm$2.68} \\
R+RFF & 673 & 96.14{\scriptsize$\pm$0.74} & 93.60{\scriptsize$\pm$47.00} & 2.78{\scriptsize$\pm$0.62} \\
ReLU & 683 & 93.01{\scriptsize$\pm$0.95} & 195.15{\scriptsize$\pm$20.10} & 5.40{\scriptsize$\pm$0.79} \\
Tanh & 683 & 88.06{\scriptsize$\pm$0.88} & 478.55{\scriptsize$\pm$67.66} & 8.85{\scriptsize$\pm$0.62} \\
IGA & 687 & 98.56{\scriptsize$\pm$0.27} & 7.64{\scriptsize$\pm$1.70} & 0.98{\scriptsize$\pm$0.18} \\
FLAIR & 687 & 96.24{\scriptsize$\pm$0.46} & 86.76{\scriptsize$\pm$28.50} & 2.81{\scriptsize$\pm$0.38} \\
TUNER & 701 & 94.62{\scriptsize$\pm$0.04} & 41.70{\scriptsize$\pm$1.75} & 3.32{\scriptsize$\pm$0.02} \\
VI3NR & 701 & 93.19{\scriptsize$\pm$0.70} & 190.38{\scriptsize$\pm$47.12} & 4.69{\scriptsize$\pm$0.47} \\
MSA-INR & 697 & 98.59{\scriptsize$\pm$0.14} & 10.56{\scriptsize$\pm$1.93} & 0.90{\scriptsize$\pm$0.11} \\
CAFE (enc.) & 709 & 93.81{\scriptsize$\pm$1.79} & 57.85{\scriptsize$\pm$26.60} & 4.59{\scriptsize$\pm$1.87} \\
CAFE & 715 & 63.62{\scriptsize$\pm$29.62} & 1430.30{\scriptsize$\pm$2061.95} & 56.88{\scriptsize$\pm$63.92} \\
\bottomrule
\end{tabular}
\end{table}

\FloatBarrier

\end{document}